\documentclass[%
 amsmath,amssymb,
 aps,
 pre
showkeys,
]{revtex4-2}

\usepackage{graphicx} %Figuras
\usepackage[utf8]{inputenc} %Para poder poner tildes
\usepackage{vmargin} %Para modificar los márgenes
\usepackage{array, makecell} %Celdas en tablas 
\usepackage{physics} %\Ln
\usepackage{amssymb} %Simbolos matemáticos 
\usepackage{stackrel} %Comando stackrel para las tasas
\usepackage{hyperref} %Hiperenlaces 
\usepackage{float} %Colocación de figuras 
\usepackage{natbib} %Bibliografía
\usepackage{xurl} %Romper las url en varias lineas y que no haya problemas al imprimir la bibliografía
\usepackage{booktabs} %Colocación de tablas
\usepackage{tabularx} %Creación de tablas
\usepackage{subfigure} %Manejo de subfiguras
\usepackage{mathtools} %Ecuaciones. Extensión de amsmath con más contenido 

\begin{document}

\title{Extinction in agent-based and collective models of bet-hedging
}

\author{Manuel Dávila-Romero$^1$}
\author{Francisco J. Cao-García$^{1,2}$}
\author{Luis Dinis$^{1,3}$}
\email{ldinis@ucm.es}
\affiliation{$^1$Departamento de Estructura de la Materia, Física Térmica y Electrónica. Facultad de Ciencias Físicas. Universidad Complutense de Madrid. Plaza de Ciencias, 1. 28040 Madrid. Spain.\\
$^2$ Instituto Madrileño de Estudios Avanzados en Nanociencia, IMDEA Nanociencia. Calle Faraday, 9. 28049 Madrid.\\
$^3$Grupo Interdisciplinar de Sistemas Complejos (GISC)}%

\begin{abstract}
Bet-hedging is a phenotype diversification strategy that combines a fast-growing vulnerable phenotype with a slow-growing resistant phenotype. In environments switching between favorable and unfavorable conditions, bet-hedging optimizes growth and reduces fluctuations over a long time, which is expected to reduce extinction risk. Here, we address directly how bet-hedging can reduce extinction probability in an agent-based model. An agent-based model is appropriate for studying extinction due to the low number of individuals close to extinction. We also show that the agent-based model converges to the collective model behavior for populations of $100$ individuals or more. However, the collective model provides relevant qualitative insight even for low populations. The collective model provides expressions for extinction that stress the relevance of the population number, 
showing that a factor four increase in the total population has a greater effect than a change of strategy from maximum growth to minimal extinction. This work provides further insight into the impact of finite population effects on the bet-hedging strategy's success. 
\end{abstract}

%Use showkeys class option if keyword display is desired

\maketitle

%\newpage
\normalsize
%%Inicio: 
\tableofcontents 
%\newpage 

%%%%% Please keep these comment symbols that help to find sections and subsections
%%%%%%%%%%%%%%%%%%%%%%%%%%%%%%%%%%%%%%%%%%%%%%%%%%%%%%%%%%%%%%%%%%%%%%%%%%%%%%
\section{Introduction}

{Biological systems often show significant levels of heterogeneity. Even in monoclonal colonies, a high degree of phenotypic variability may be present. Different amounts of variability have been found in many measurable traits in microbial populations, such as individual growth rate \cite{bigger_treatment_1944}, cell size \cite{lin_effects_2017}, doubling times \cite{pugatch_greedy_2015}, asymmetries in protein amounts after division  \cite{bergmiller_biased_2017} and others. Whether this phenotypic variability is just a consequence of fluctuations and noise in the processes 
expressing the genotype imparing fitness or it is instead a trait favored by natural selection is still an open question \cite{levien_non-genetic_2021,eldar_functional_2010}. For instance, phenotypic diversity can be beneficial through the protection it confers to face uncertainty through bet-hedging. Bet-hedging in biological systems consists of diversifying strategies, i.e., phenotypes, seeking protection against random changes in environmental quality. 
A bet-hedging strategy sacrifices the growth of some individuals during a favorable environment to improve population survival odds in the event of rapidly deteriorating environmental conditions. Typical examples show the coexistence of a fast-growing but vulnerable phenotype with a slow-growth tolerant, persistent, or dormant phenotype 
%and have been widely studied from a biological perspective 
\cite{levin_dormancy_2013,morawska_diversity_2022, lennon_microbial_2011}. } Bet-hedging also showcases the importance of epigenetic inheritance in cellular differentiation \cite{veening_bet-hedging_2008}. Delayed germination in semi-desert plants \cite{venable_bet_2007} or the appearance of antibiotic-resistant cells with a low growth rate in favorable conditions \cite{morales_targeting_2022}, are interpreted as examples of bet-hedging against environmental fluctuations in nature. 

{Additionally, bet-hedging can be effective against demographic noise that poses an extinction risk to small populations \cite{xue_bet_2017}. A colony might maximize its future abundance by diversifying into coexisting phenotypes of ``fast-growers'' and ``better-survivors'' even in a constant environment under certain conditions.} Hedging is also relevant in finance as a risk control strategy \cite{tse_recent_2002}, and connections between economic principles and cell behavior have been analyzed in detail in Ref. \cite{collective_economic_2023}.

{Here, we consider a model for population growth in a changing environment featuring bet-hedging strategies in the form of coexisting phenotypes, as in Ref. \cite{hufton_phenotypic_2018}.
We consider the simplest choice of two possible environments and two possible phenotypes.} The individuals switch between the two phenotypes with tunable rates while the environment switching rate is fixed, and the performance of the phenotypes in each of the two environments is also fixed. {The model is simple but well suited, for instance, to describe exponential growth in bacterial colonies subjected to changing environmental conditions.}

This dynamics has been previously addressed by Dinis et al. in Ref.~\cite{dinis_pareto-optimal_2022} in the large population limit, using a {deterministic piece-wise Markov process where each phenotype is treated as a uniform population, which we will refer to as a} collective approximation {or collective model}. Here, we add demographic fluctuations in the kinetics via an agent-based model.
Ref.~\cite{dinis_pareto-optimal_2022} addressed the question of the optimal bet-hedging strategy for a population in an environment switching between favorable and unfavorable conditions. The aim was to maximize the growth and minimize the variance. The optimal switching rates between phenotypes differ depending on the relative weight of those two objectives.
The set of all the optimal points with different relative weights is called the Pareto front, {shown in Fig. \ref{fig:ParetoExtinctionVSalpha}(a), in terms of the intensity of the fluctuations versus the growth rate.
The rightmost point in the Pareto front corresponds to the maximum attainable growth rate of the colony. The slope of the front is extremely large at that point, implying that a slight reduction in growth rate allows a beneficial large reduction of fluctuations and extinction risk.} 

Here, we compute the extinction probability for these bet-hedging models. 
The proper study of extinction requires {considering a low population.  A collective model cannot describe precisely this limit, whereas an agent-based model, like the one we develop in this paper, can. Our model thus combines both environmental and demographic fluctuations}. The results are also compared here with the results expected with extrapolation of the collective model, which provide useful simple approximations. 

In Section \ref{sec:models}, we present the models and approximations. We describe the agent-based model and the collective mode approximation with a summary of the main results for this approximation of Ref. \cite{dinis_pareto-optimal_2022}.  
In Section \ref{sec:results}, we show the trajectories of the agent-base model for short times and long times, comparing the long-time behavior of the total number of individuals with the results from the collective model. In Section \ref{sec:extinction}, we present the extinction analysis for the agent-base model, and their results are compared with the collective mode results. Finally, we discuss and comment on the results.

%%%%%%%%%%%%%%%%%%%%%%%%%%%%%%%%%%%%%%%%%%%%%%%%%%%%%%%%%%%%%%%%%%%%%%%%%%%%%%%%%%
\section{Bet-hedging model} \label{sec:models}

We study here a two-phenotype bet-hedging model, where the phenotype switching rate is independent of the environment. 
We first present an agent-based model described in Sec. \ref{sec:agentmodel}, which addresses individuals' dynamics, making it a model particularly appropriate for low-population dynamics and extinction computations. We then present in Sec. \ref{sec:collectivemodel} a collective model valid for large populations, which provides a master equation for the population number in each phenotype. Predictions of both models for growth and extinction are then compared in the Results Section, Sec. \ref{sec:results}.

%%%%%%%%%%%%%%%%%%%%%%%%%%%%%%%%%%%%%%%%%%%%%%%
\subsection{Agent-based model} \label{modelodiscreto} \label{sec:agentmodel}

In the agent-based model, each individual follows the dynamics described in Fig.~\ref{fig:populationdynamics}.
We consider stochastic dynamics for individuals of a population in two possible environments, $E_0$ (good environment) and $E_1$ (bad environment), and two phenotypes, each suitable for a different environment, $A$ (fast-growing {and sensitive}) and $B$ (slow growing resistant). Environments randomly switch with rates $\kappa_{0\to 1}$ and $\kappa_{1\to 0}$, 
    \begin{equation*}
        E_0 \underset{\kappa_{0\to 1}}{\stackrel{\kappa_{1\to 0}}{\rightleftarrows}} E_1.
    \end{equation*}
We consider here the case where both environments are equally probable, $\kappa_{0\to 1} = \kappa_{1\to 0}$, and use its characteristic time as time-scale, $\kappa_{0\to 1} = \kappa_{1\to 0}=1$.

Each phenotype has a characteristic growth rate in each environment. Phenotype $A$ has a fast-growing sensitive profile, with a growth rate of $k_{A0} = 2$ in the good environment and $k_{A1}=-2$ in the bad environment. Instead, Phenotype $B$ has a slow-growing resistant profile, with growth rates an order of magnitude smaller, $k_{B0}=0.2$ and $k_{B1}=-0.2$. 
{
We consider all individuals of the same phenotype equal, even after reproduction (a simplification that applies to bacteria, for example).}

Phenotypes of each individual switch randomly with rates, $\pi_{A\to B}$ and $\pi_{B\to A}$,
    \begin{equation*}
        A \underset{\pi_{A\to B}}{\stackrel{\pi_{B\to A}}{\rightleftarrows}} B.
    \end{equation*}
These switching rates, $\pi_{A\to B}$ and $\pi_{B\to A}$, are the {parameters to be optimised.}

We consider continuous-time dynamics, which requires that the time steps in the numerical implementation are smaller than all the characteristic time scales, given here by the inverses of the rates. We consider an initial population of $5$ individuals in each phenotype, $n_A=5$ and $n_B=5$. Thus, the total initial population is $n=n_A+n_B=10$ individuals, a low population allowing us to see extinction effects.

\textbf{\begin{figure}%[H] 
 \centering
\includegraphics[draft=false,scale=0.5]{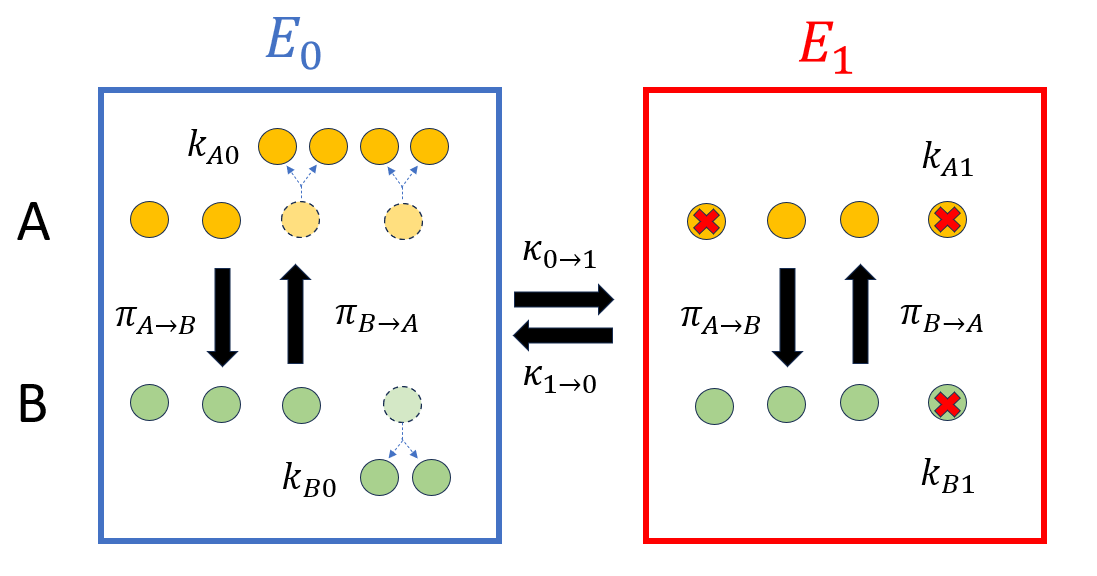}
\caption{\textbf{Outline of the population dynamics.} The population presents two phenotypes $A$ and $B$. Individuals switch between the phenotypes with rates $\pi_{A\to B}$ and $\pi_{B\to A}$. $A$ is more sensitive to the environment. In Environment $E_0$ both phenotypes reproduce with individual rates $ k_{A0} > k_{B0} > 0 $, while in Environment $E_1$ both phenotypes die with individual rates $ k_{A1} < k_{B1} < 0 $. Environments randomly change with rates $ \kappa_{0\to 1} $ and $ \kappa_{1\to 0} $. The question we address is: with all the other rates fixed ($\kappa_{0\to 1}=\kappa_{1\to 0}=1; \,\,k_{A0}=2,\,k_{B0}=0{.}2,\,k_{A1}=-2,\,k_{B1}=-0{.}2$),
which are the optimal values of the phenotype switching rates $\pi_{A\to B}$ and $\pi_{B\to A}$? This also requires to define what we consider optimal, as discussed in Sec.~\ref{sec:collectivemodel}. (As initial conditions, we consider a population of $10$ individuals, $5$ in each phenotype.) }
\label{algoritmo}\label{fig:populationdynamics}
\end{figure}}

%\subsection{Binomial}
In the agent-based model, each process (change of environment, reproduction or death of individuals, and phenotypic change) is stochastic. A random number $r$ with a uniform distribution between $0$ and $1$ is generated to decide whether the process occurs. The process occurs if and only if $r < a \Delta t $, with $a$ the rate of the process and $\Delta t$ the numerical discretization time step. 
At each time step of the simulation, a random number for the change of environment is generated first. After that, the total population of $N$ bacteria {should be} iterated through, and for each of them, another random number generated to decide whether it reproduces (for positive growth rate $k>0$) or dies (for negative growth rate $k<0$). {Here we take a slightly different approach. In principle a random number would be needed for each existing individual to see if its phenotype changes}. In other words,  $2N+1$ random numbers {would be} required: one for the environment, $N$ for reproduction/death, and $N$ for the phenotype change. When reproduction is a much more common phenomenon in the dynamics than death, the growth rate can become quite high. This results in many individuals and many random numbers generated, making the program inefficient. Since only one random number is needed for the environment, we will optimise the other two processes (growth and phenotype change) {in the following manner}.

Phenotype change and growth have only two possible outcomes: occurring with a certain probability $p$ or not occurring with probability $q=1-p$. This is precisely the definition of a random experiment of the Bernoulli type \cite{noauthor_bernoulli_2008}. If the experiment is repeated for the $N$ bacteria, the Bernoulli distribution gives rise to the {binomial distribution} \cite{noauthor_binomial_2008}. The binomial distribution is the discrete probability distribution that counts the number of successes in a sequence of $n$ independent Bernoulli trials with a fixed probability of occurrence. In this distribution, we associate the new probability $p$ with the quantity $a \Delta t$, where $a$ is the rate of the process analyzed.
Thus, in each step, only a random number following a uniform distribution for the environment change and {four} random numbers distributed according to a binomial distribution ({two} for reproduction/death and two for the phenotype change since we evolve the populations of phenotypes $A$ and $B$ separately) need to be generated. 
{
This binomial approach is exact for every number of individuals, not an approximation, as all the individuals of each phenotype are equal, and we only need the change in the number of individuals in the phenotype. The binomial approach }
allows for a much more efficient and faster program. The random numbers with uniform or binomial distribution are generated using the \textit{random} package of Python \cite{van1995python}. The code used for these simulations can be found at Ref. \cite{codigo}.

%%%%%%%%%%%%%%%%%%%%%%%%%%%%%%%%%%%%%%%%%%%%%%%%%%%%%%
\subsection{Collective model} \label{sec:collectivemodel}
\label{bet hedging ambientes fluctuantes}

When the population is large (or the order of $100$ individuals or more) we can adopt the approach of Ref. \cite{dinis_pareto-optimal_2022}, and consider a collective model description providing a master equation for the population number of both phenotypes at each environment.
Let $X(t)=(n_A(t),n_B(t))^T$ be the population vector that describes the number of individuals of each phenotype in an instant $t$. The vectorial master equation is 
\begin{equation} 
    \frac{d}{dt}X(t)=M_{E(t)}X(t)
    \label{ecuaciondiferencial}
\end{equation}
with $E(t)$ the environment at time $t$, with values either $E_0$ (good environment) or $E_1$ (bad environment), which have the transition matrices
\begin{equation*}
M_{E_0}=\left(\begin{array}{cc}
k_{A 0}-\pi_{A\to B} & \pi_{B\to A} \\
\pi_{A\to B} & k_{B 0}-\pi_{B\to A}
\end{array}\right); \quad \quad M_{E_1}=\left(\begin{array}{cc}
k_{A 1}-\pi_{A\to B} & \pi_{B\to A} \\
\pi_{A\to B} & k_{B 1}-\pi_{B\to A}
\end{array}\right).
\end{equation*}

For the total population $n(t)=n_A(t)+n_B(t)$ {at finite time $t$}, we can define the average effective growth rate as  
\begin{equation} \label{tasacrecimientofinita}
    \Lambda_t=\frac{1}{t}\ln\frac{n(t)}{n(0)},
\end{equation}
with a long time value of 
\begin{equation}
    \Lambda=\lim_{t\to\infty} \Lambda_t.    
\end{equation}
(Both $\Lambda_t$ and $\Lambda$ have dimensions of the inverse of time.)

The average effective variance is defined as
\begin{equation}
    \text{Var}(\Lambda_t)= \left<\left(\Lambda_t\right)^2\right>-\left< \Lambda_t \right>^2 ,
\end{equation}
{where the average is taken over different realizations of the stochastic environmental trajectories. This variance of the finite-time growth rate}
 is inversely proportional to $t$ asymptotically, $ \lim_{t\to\infty}  \text{Var}(\Lambda_t) \sim 1/t$, according to the Central Limit Theorem \cite{noauthor_central_2008}. The asymptotic proportionality constant of the variance of the growth rate could be defined (abusing notation) as $ \text{Var}(\Lambda) = \lim_{t\to\infty} t \text{Var}(\Lambda_t)$, as in Ref.~\cite{dinis_pareto-optimal_2022}. (Thus, $\text{Var}(\Lambda_t)$ has units of the inverse of time squared, while $\text{Var}(\Lambda)$ has units of the inverse of time.) 
{
However, we prefer to express the limit as
\begin{equation}
    \lim_{t\to\infty}  \text{Var}(\Lambda_t) = \frac{2D}{t},
\end{equation}
where $D$ can be interpreted as an effective diffusion constant for the dispersion of the logarithm of the total population number, as we show later in Sec.~\ref{sec:long-term}. The relation between the two constants is $\text{Var}(\Lambda)=2D$.
}

Thus, the variance of the growth decreases with time, but at finite time the question of balancing the growth and its variance is present, mainly motivated by the extinction risk. How to tune the switching rates $\pi =(\pi_{A\to B},\pi_{B\to A})$
to balance the maximization of the growth and the minimization of the variance 
was already addressed in the collective model in Ref. \cite{dinis_pareto-optimal_2022}. The authors of that work proposed to minimize the objective function $J(\pi) = -\alpha\Lambda(\pi) + (1-\alpha)\sqrt{\text{Var}(\Lambda(\pi))}$, creating a Pareto front.
It was shown that moving along the Pareto front, we can significantly reduce the variance with a small penalty in growth. {In the notation chosen here, this objective function is $J(\pi) = -\alpha\Lambda(\pi) + (1-\alpha)\sqrt{2D(\pi)}$. }

\begin{figure}%[H] 
    \centering
    \subfigure[]{\includegraphics[draft=false,width=0.45\textwidth]{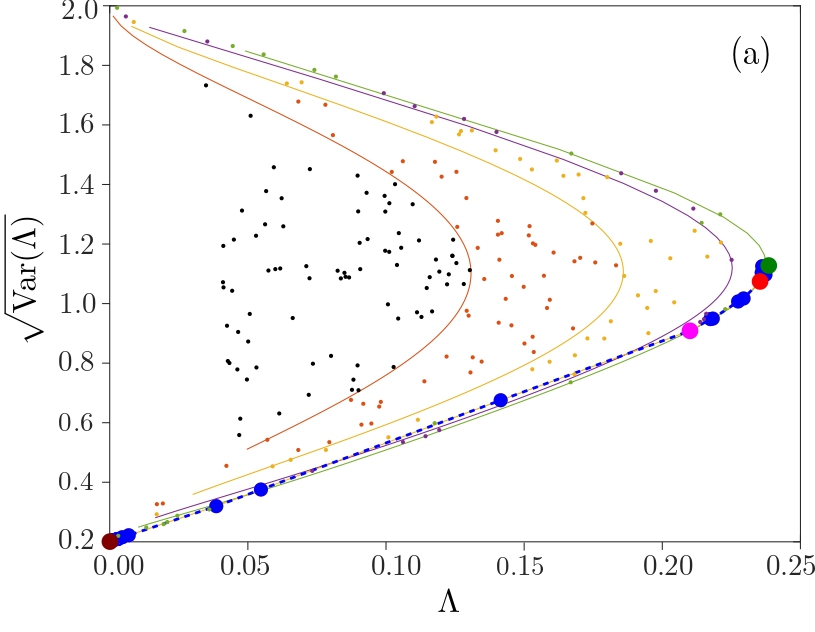}} 
    \subfigure[]{\includegraphics[draft=false,width=0.45\textwidth]{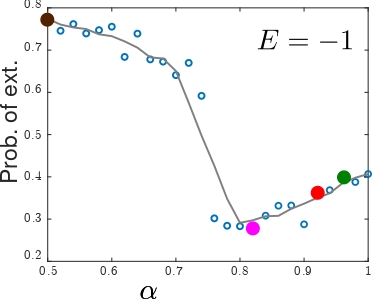}} 
    \caption{\textbf{Pareto front and probability of extinction on the Pareto front for the collective model} introduced in Sec.~\ref{sec:collectivemodel} and in Ref.~\cite{dinis_pareto-optimal_2022}. Panel (b) shows that the maximum growth case (green point) does not give the minimum extinction. This is due to its larger fluctuations. {(Panels (a) and (b), figures originally appeared in Ref.~\cite{dinis_pareto-optimal_2022}, as a courtesy of the authors).
    }
    }
    \label{figuras 1.2}
    \label{fig:ParetoExtinctionVSalpha}
\end{figure}

For the values of the parameters given in Sec.~\ref{sec:agentmodel} and Fig.~\ref{fig:populationdynamics}, the Pareto front has been found (in Ref.~\cite{dinis_pareto-optimal_2022}) to be approximately parameterized by 
\begin{equation} \label{eq:Pareto-front-expression}
    \frac{1}{2}\left(\frac{1}{\pi_{A\to B}}+\frac{1}{\pi_{B\to A}}\right)\approx T=3{.}33.
\end{equation}
Fig.~\ref{fig:ParetoExtinctionVSalpha} shows the Pareto front and the extinction values for the switching rates at the Pareto front obtained in Ref.~\cite{dinis_pareto-optimal_2022}. The result stresses that the minimum extinction is not reached at the maximum growth, due to the increased associated fluctuations for these case; see Table~\ref{tab:comparison}. This table also shows the optimal switching rates for the maximum growth (second column), and for the minimum variance (last column) cases. Note that just minimizing the variance leads to a very low growth. These previous results raised the question of to what extend we need to minimize the variance to minimize extinction {when demographic fluctuations are present, which we address in this manuscript.}

%%%%%%%%%%%%%%%%%%%%%%%%%%%%%%%%%%%%%%%%%%%%%%%%%%%%%%%%%%%%%%%%%%%%%%%
\section{Results} \label{sec:results}\label{resultados}

{To better grasp the dynamics of the system, we} first study the trajectories in the agent-based model, both in the short-time and in the long-time limits, comparing long-time evolution with the collective model. Second, we address the computation of extinction probability with the agent-based model and compare the results with the collective model. Finally, the probability density functions of the time of extinction obtained from simulations of the agent-based model are compared with the predictions from the collective model. 

%%%%%%%%%%%%%%%%%%%%%%%%%%%%%%%%%%%
\subsection{Trajectories of the agent-based model} \label{sec:traj-agent-based}

In this section, the population trajectories in the agent-based model are analyzed, establishing a comparison with the collective model. This comparison highlights the validity regime of the collective model as an approximation to the long-time behaviour of the agent-based model.

%%%%%%%%%%%%%%
\subsubsection{Short-term evolution: Trajectories of the population of each phenotype} \label{sec:short-term}

At short times, when we start with a low population (as in Fig.~\ref{fig:short-time-agent} where initial conditions are of $10$ individuals, $5$ of each phenotype) the discrete population effects are particularly noticeable for the fast growth sensitive phenotype $A$. They become less important as the population reaches {about} $100$ individuals ($\ln(100)=4.6$), starting the medium-time evolution.

At short and medium times, the effects of the environment switching are visible. Fig.~\ref{fig:short-time-agent} shows in Panel (a) that the logarithm of the population of the phenotypes, $\ln n_A(t)$ and $\ln n_B(t)$, are constrained to a region {of the plot}. It can be shown that the limits of this region approximately correspond to the respective equilibrium balance between phenotypes at each environment. 
These equilibrium conditions were previously obtained in the collective model in Ref.~\cite{dinis_pareto-optimal_2022}. 
Let $\Delta_\sigma=k_{A\sigma}-k_{B\sigma}$ where $\sigma=0,1$ indicates the value in the environment $E_\sigma$ ($\sigma = 0 $ denotes the good environment $E_0$ and $\sigma =1 $ the bad environment $E_1$). Let also $\pi_T=\pi_{A\to B}+\pi_{B\to A}$. For the collective model \cite{dinis_pareto-optimal_2022}, it can be proven that, given an environment, the relative fraction of phenotype $A$, defined as $\phi=\frac{n_A}{n_A+n_B}$, evolves independently of the population magnitude. An analysis of $\phi(t)$ shows that, for each environment, there are two fixed points given by
\begin{equation} \label{phi}
    \phi^{\pm}_\sigma=\frac{\Delta_\sigma-\pi_T\pm\sqrt{(\Delta_\sigma-\pi_T)^2+4\pi_{B\to A}\Delta_\sigma}}{2\Delta_\sigma},
\end{equation}
where $\phi^-$ is an unstable fixed point and $\phi^+$ a stable one.
Now, the following relationship between the populations of the two phenotypes can be written as a function of $\phi$
\begin{equation} 
    \frac{n_B}{n_A}=\frac{1-\phi}{\phi}=\frac{1}{\phi}-1.
\end{equation}
Using logarithm in this expression
\begin{equation} \label{log equilibiro}
    \ln\left(\frac{n_B}{n_A}\right)=\ln n_B-\ln n_A=\ln\left(\frac{1}{\phi}-1\right),
\end{equation}
which can be specialized for the $\phi^{+}$ point of each environment. When plotting $\ln n_B(t)$ against $\ln n_A(t)$, as in Fig.~\ref{fig:short-time-agent}, Eq. \eqref{log equilibiro} is a line with unit slope and environment dependent intercept. Since it is given by a stable equilibrium point of $\phi$, this line is an attractor for the evolution. There are two attractor lines, one for each environment.

Analyzing the trajectories (see Fig.~\ref{fig:short-time-agent}), we see that the population effectively oscillates between the two attractor lines, approaching one or the other depending on the environment. Furthermore, we see that the evolution in the bad environment $E_1$ is dominated by a reduction of the population of phenotype $A$, while the population of phenotype $B$ remains approximately constant. We also see that the population grows on average, mainly thanks to events of short bad environment intervals followed by long good environment intervals. When the opposite happens, the long bad interval shifts the population to phenotype $B$, protecting the population but stalling growth until a future long good interval turns the tables.
These processes lead to the alternating stall and growth behavior observed for the total population evolution; see Panel (d) of Fig.~\ref{fig:short-time-agent}.
    
\textbf{\begin{figure}%[H] 
 \centering
\includegraphics[draft=false,scale=0.7]{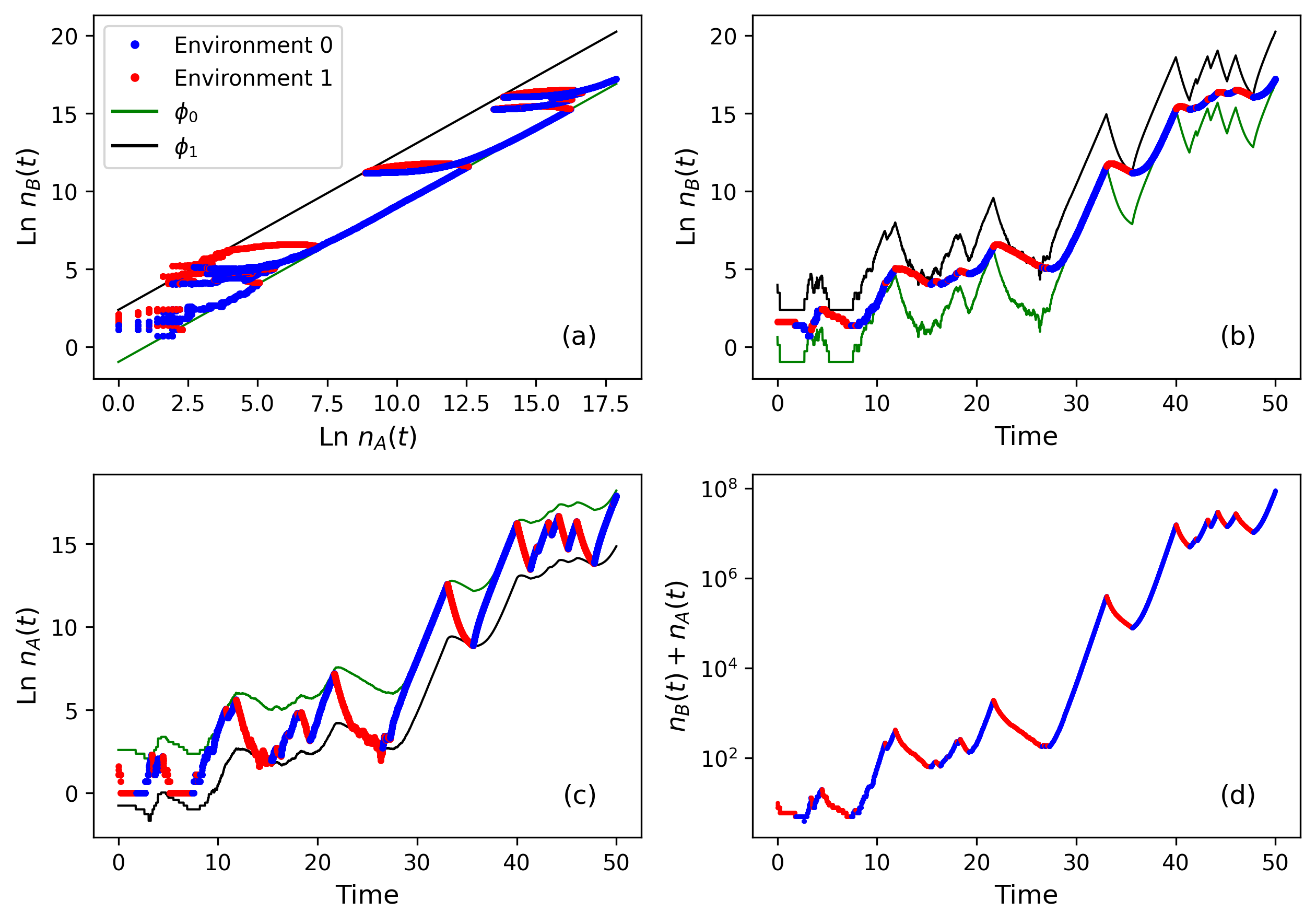}
\caption{\textbf{Short and medium time trajectory for minimum extinction in the Pareto front.} {At short times, the population is low and discreteness is apparent in the evolution, reflected in the trajectory as a collection of separate points. Simulations are run} with phenotype switching rates given in Table~\ref{tab:comparison}, values of the other rates given in the caption of Fig.~\ref{fig:populationdynamics} and initial total population $10$ individuals ($5$ of each phenotype).  Panel (a) shows how the trajectory in the plane $(\ln n_A(t),\ln n_B(t))$ is constrained between the equilibrium lines of the environments, $\phi_0$ and $\phi_1$. The other panels depict the evolution of the logarithm of the populations of phenotypes and of the total population over time.}
\label{fig:short-time-agent}
\end{figure}
}

%%%%%%%%%%%%%%%%%%%%%
\subsubsection{Long-term evolution: Trajectories of the total population} \label{sec:long-term}

\begin{figure}%[H] 
    \centering
    \subfigure[]{\includegraphics[draft=false,width=0.48\textwidth]{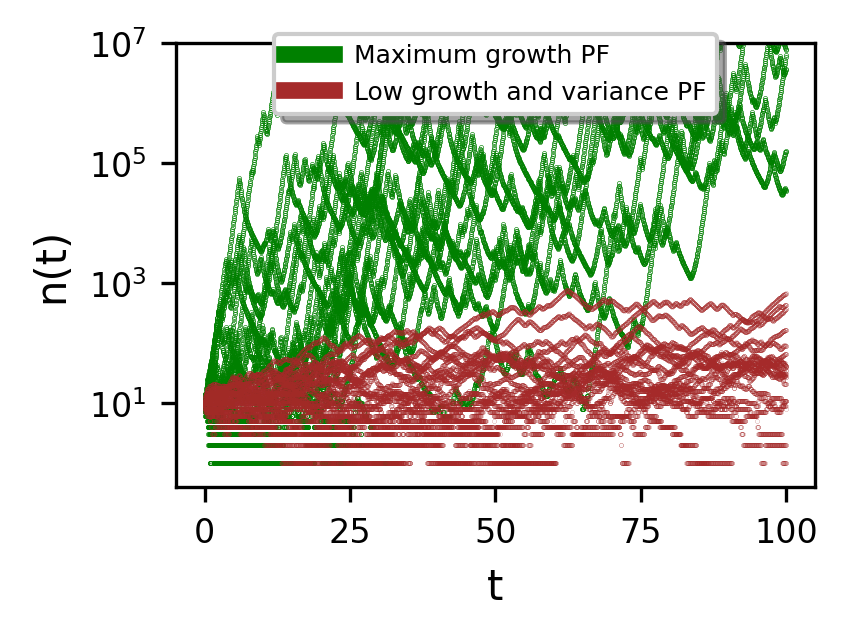}
    \label{picture label 2.a}}
    \subfigure[]{\includegraphics[draft=false,width=0.48\textwidth]{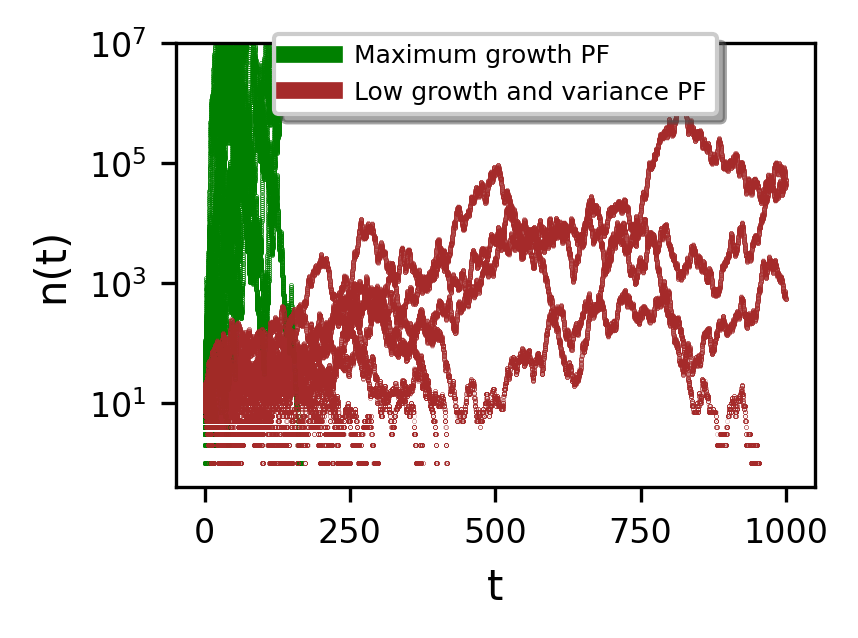}
    \label{2.b}}

    \caption{\textbf{Long time trajectories for the maximum growth in the Pareto front and the low growth and variance in the Pareto front} computed with the agent-based model (switching rates are given in first and last rows of Table~\ref{tab:comparison}, respectively, the other rates and the initial conditions are described in caption of Fig.~\ref{fig:populationdynamics}). 30 simulations are represented for each case.
    }
 %   \label{trayectorias completas}
 %   \label{trayectorias completas 2}
    \label{fig:long-time-agent}
\end{figure}

\begin{figure}%[H] 
    \centering
    \subfigure[]{\includegraphics[draft=false,width=0.48\textwidth]{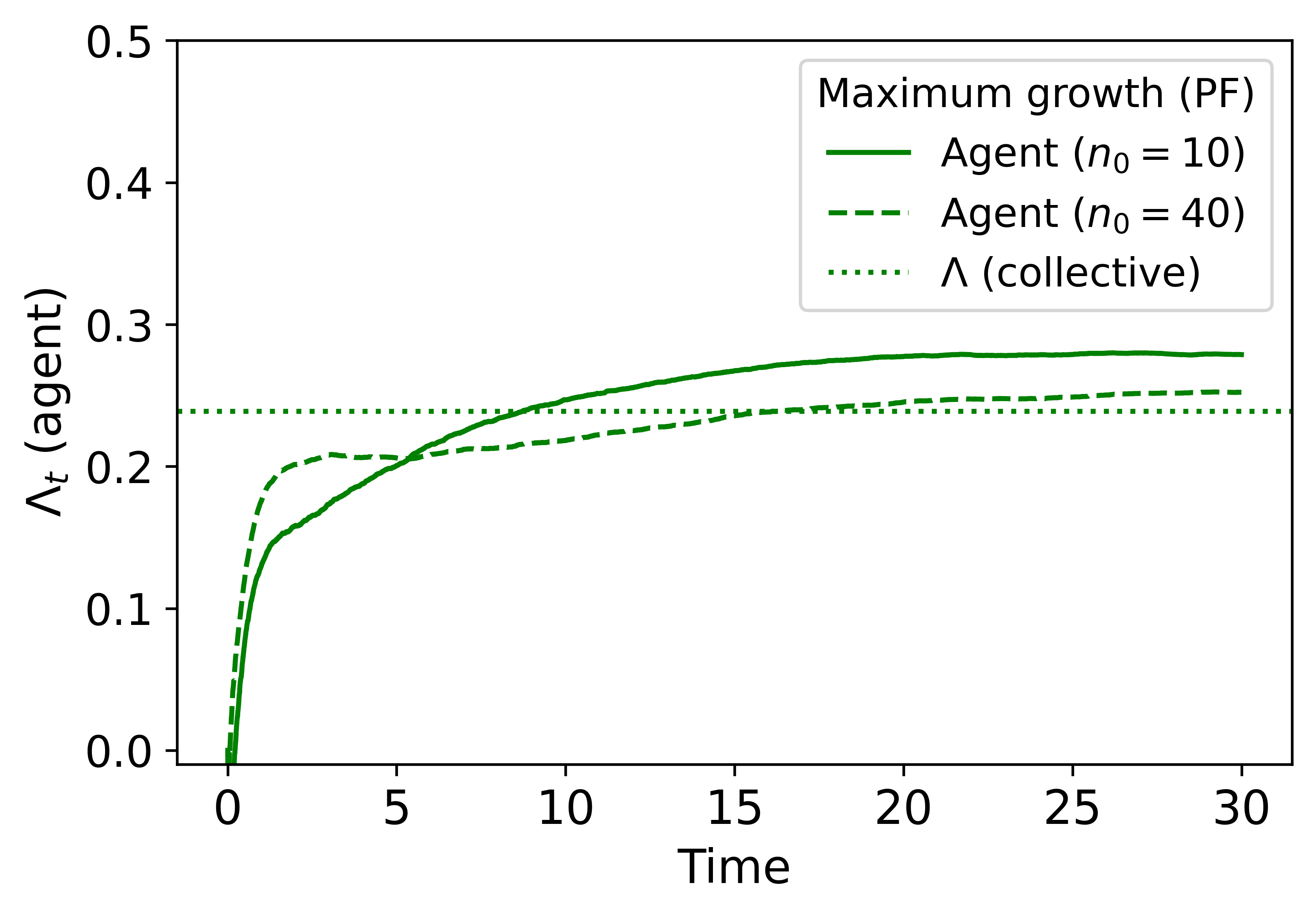}
    \label{picture label 5.c}}
    \subfigure[]{\includegraphics[draft=false,width=0.48\textwidth]{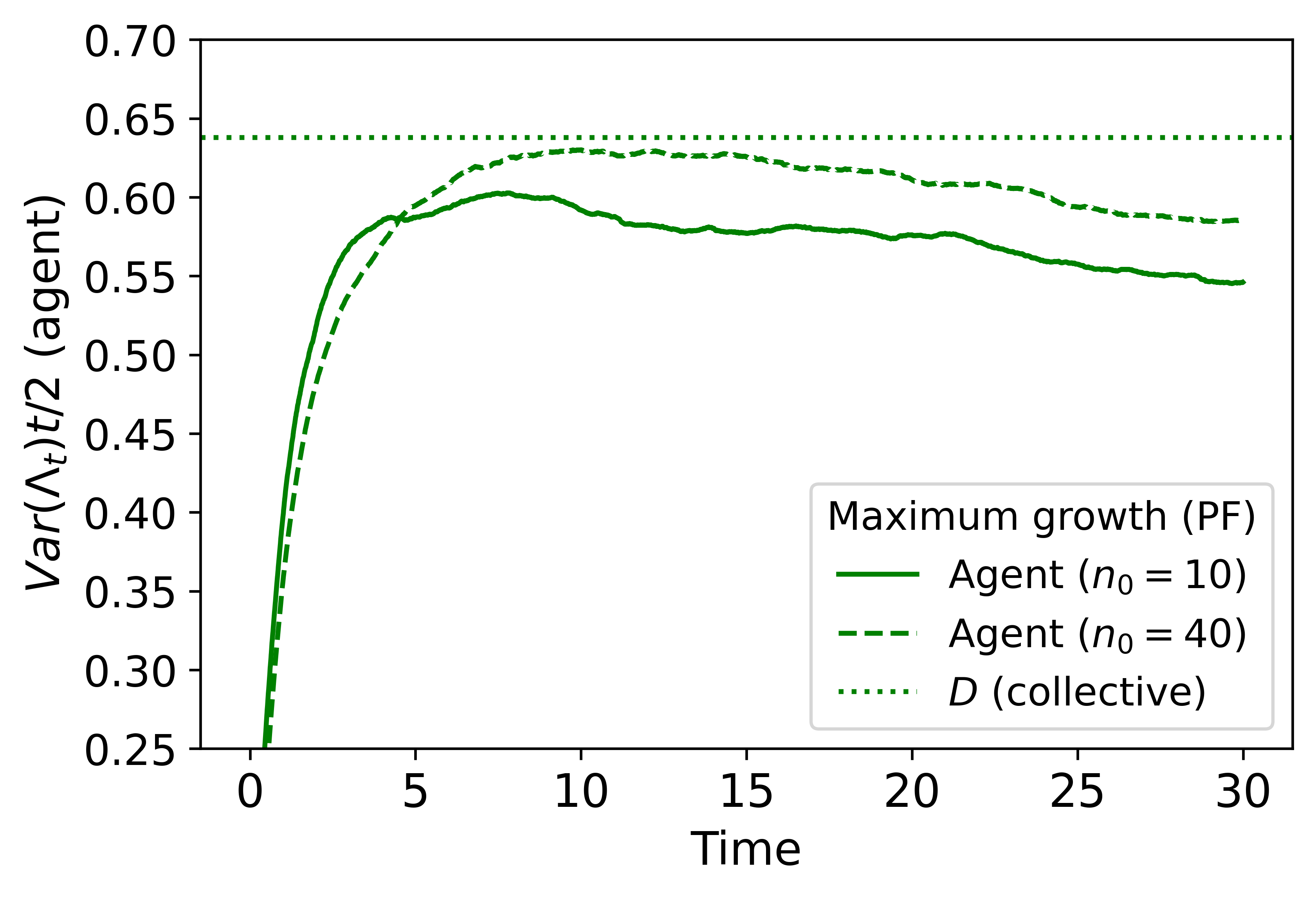}
    \label{5.d}}
    \subfigure[]{\includegraphics[draft=false,width=0.48\textwidth]{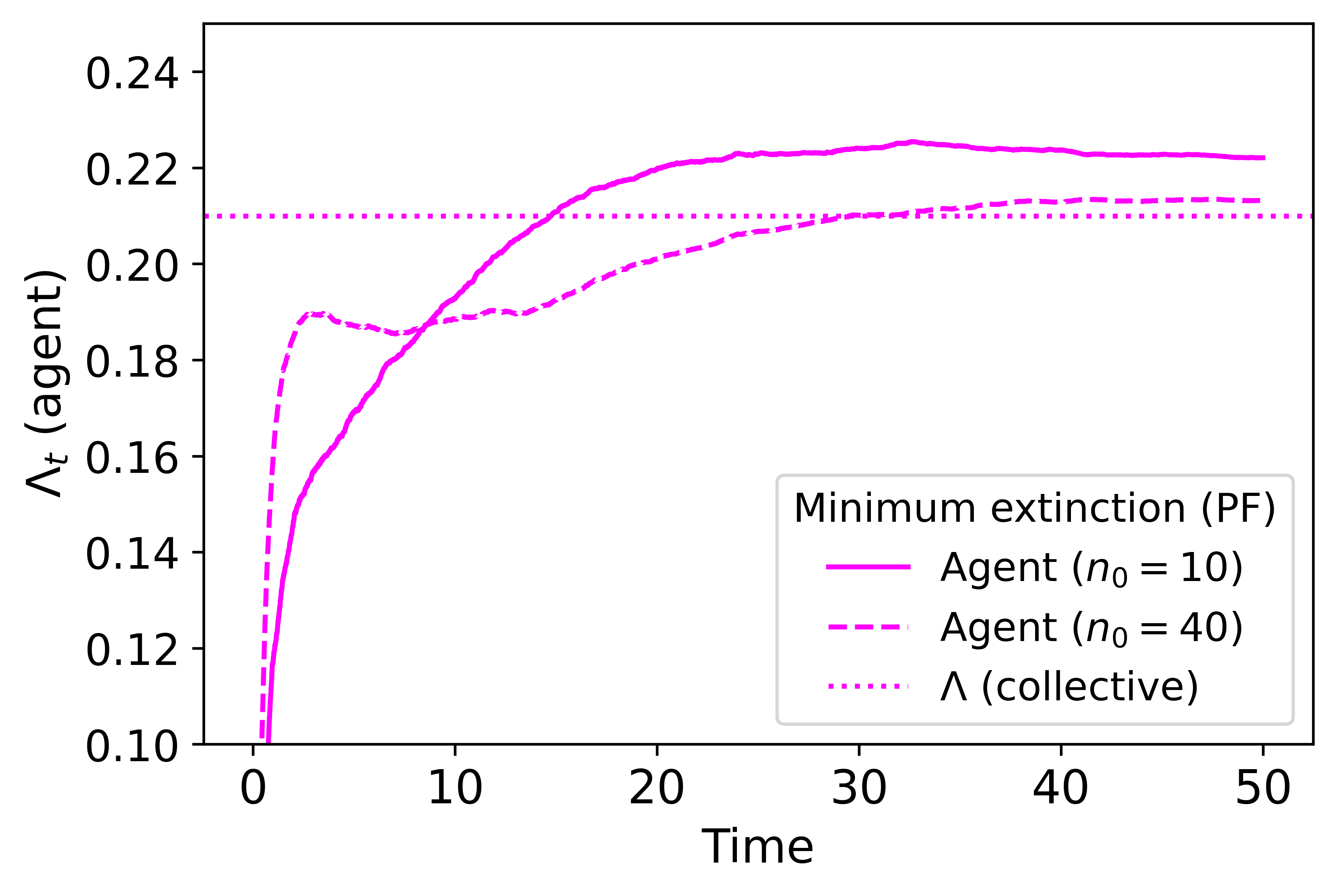}
    \label{picture label 5.a}}
    \subfigure[]{\includegraphics[draft=false,width=0.48\textwidth]{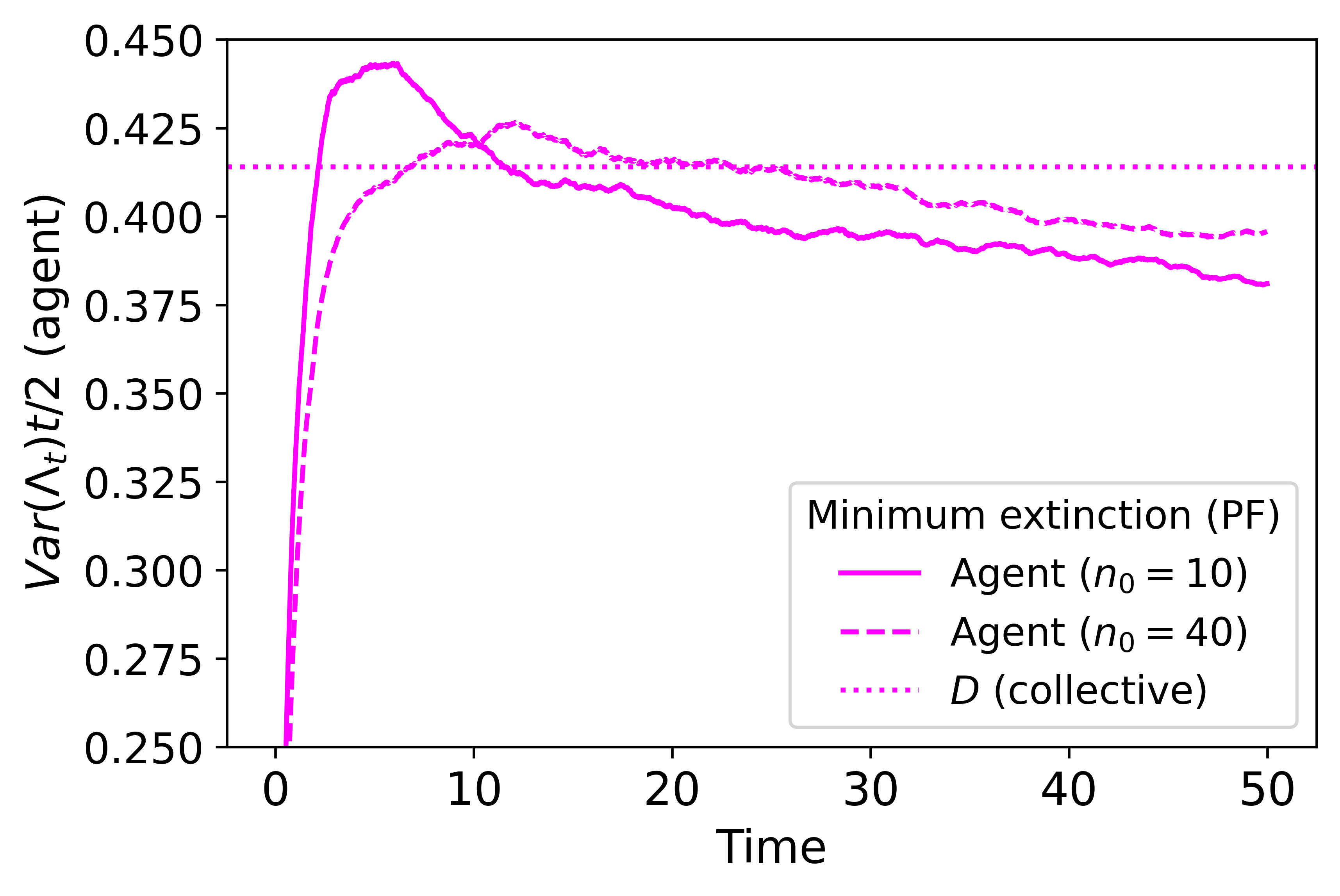}
    \label{5.b}}

    \caption{\textbf{Growth and diffusion time evolution computed with the agent-based model.} 
    The panels show $\left< \Lambda_t\right>$ and $ \text{Var}(\Lambda_t) t/2$, estimators of $\Lambda$ and $D$, as defined in Sec.~\ref{sec:collectivemodel}, computed with the trajectories of the surviving populations. %The asymptotic value reached is shown in Table~\ref{tab:comparison}. 
    The dotted lines correspond to the analytical values from the collective model with the same phenotypic switching rates, shown in Table~\ref{tab:comparison}. 
    $10^4$ simulations were used to compute the expectation values.
    (Switching rates are given in the second row of Table~\ref{tab:comparison}, the other rates and the initial conditions are described in the caption of Fig.~\ref{fig:populationdynamics}.) 
    } 
    \label{fig:mean-variance}
\end{figure}

We consider long-times times much greater than the characteristic times in each environment (\textit{i.e.}, $ t \gg 1/\kappa_{0\to1} + 1/\kappa_{1\to0}$.
The stall and growth behavior at medium times (see Panel (d) of Fig.~\ref{fig:short-time-agent}) becomes, at long times, a sustained growth with fluctuations (see Fig.~\ref{fig:long-time-agent}), with approximately constant growth rate and diffusion constant, defined in Sec.~\ref{sec:collectivemodel}. Fig.~\ref{fig:mean-variance} shows that for long times, the growth rate and its diffusion constant stabilize around their asymptotic values. 
Table~\ref{tab:comparison} compares the asymptotic values obtained with the agent-based model simulations to the collective model's predictions. They show a reasonable agreement except for the low growth and variance case, where the collective model underestimates the growth rate.

\begin{figure}%[H]
    \centering
    \subfigure[]{\includegraphics[draft=false,width=0.45\textwidth]{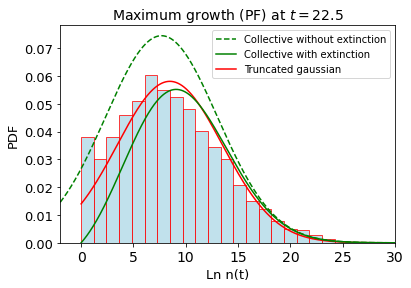}} 
    \subfigure[]{\includegraphics[draft=false,width=0.45\textwidth]{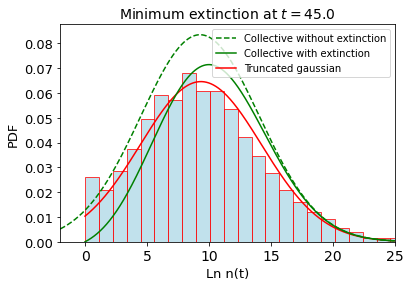}} 
    \subfigure[]{\includegraphics[draft=false,width=0.45\textwidth]{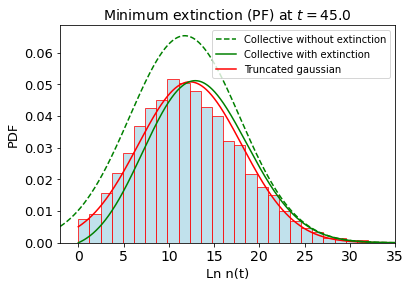}} 
    \subfigure[]{\includegraphics[draft=false,width=0.45\textwidth]{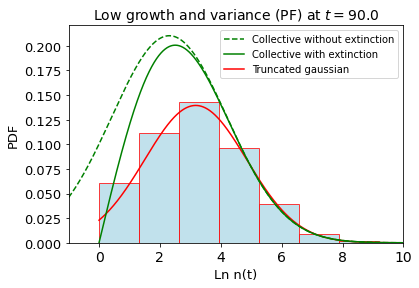}}
    \caption{\textbf{Probability density function (PDF) for the logarithm of the total population.} Bars are computed using $10000$ simulations of the agent-based model. They are compared to a truncated Gaussian with mean and standard deviation equal to those of the sample of the simulations and to the collective model predictions.
    The four panels represent the cases described in Table~\ref{tab:comparison} where agent-based and collective results for the growth and its variance are compared.  
    (The other rates and the initial conditions are given in caption of Fig.~\ref{fig:populationdynamics}.)
    In dashed green, the collective \ref{modelodiscreto} behaviour using Eq. \eqref{eq:pure_gaussian} approximation is represented. In the solid green line, the collective model approximation Gaussian is depicted corrected with absorbing boundary conditions at $\ln(n)=0$, i.e., $n=1$. See Eq. \eqref{gaussinaconextincion}. In red, the truncated Gaussian with mean and standard deviation equal to those of the sample of the simulations and same area as the simulated histogram is represented.
    }
    \label{fig:pdf-comparison}
    \label{gaussianotocho}
\end{figure}

The collective model describes the dynamics in terms of the growth rate $\Lambda_t \xrightarrow[]{t\to\infty} \Lambda $ and its variance $ \text{Var}(\Lambda_t) \xrightarrow[]{t\to\infty} \text{Var}(\Lambda) / t = 2D / t $. We can associate with this description a Gaussian distribution for the probability density function of the logarithm of the total population $x=\ln(n)$ at a given late enough time. 
The mean of $x$ is given by 
\begin{equation} \label{eq:mu}
    \mu(t) = \left< x(t)\right> = \left< x_0+x(t)-x_0\right> = x_0 + \left< \ln(n(t)/n(0))\right> = x_0 + \left< \Lambda_t\right> t   
    \xrightarrow[]{t\to\infty} x_0 + \Lambda t,   
\end{equation}
where $x_0=x(0)=\ln(n(0))$ is the logarithm of the initial value of the total population. The variance of $x$ is given by
\begin{equation} \label{eq:sigma2}
\begin{split}
  \sigma^2(t) &= \left< (x(t)-\mu(t))^2\right> = \left< (x(t)-x(0)-\left< \Lambda_t\right> t)^2\right> = \left< (x(t)-x(0)-\left< \Lambda_t\right> t)^2\right> = \\ 
  &= \left<  (\Lambda_t t - \left< \Lambda_t\right> t)^2 \right> = \text{Var}(\Lambda_t) t^2 \xrightarrow[]{t\to\infty} 2 D t
\end{split}
\end{equation}
where $ D = \text{Var}(\Lambda) / 2 = \lim_{t\to\infty} t \text{Var}(\Lambda_t) / 2 $. We therefore have the following Gaussian estimate for the probability density function (PDF) for the value of $x$ (the logarithm of the total population) at time $t$
\begin{equation} \label{eq:pure_gaussian}
    p(x, t|x_0)=\frac{1}{\sqrt{2 \pi \sigma^2(t)} }\exp \left[\frac{-\left(x-\mu(t)\right)^2}{2 \sigma^2(t)}\right]
    =\frac{1}{\sqrt{4 \pi D t} }\exp \left[\frac{-\left(x-x_0-\Lambda t\right)^2}{4 D t}\right].
\end{equation}
This PDF is expected to provide a good approximation for late times, provided extinction can be neglected (as it does not consider the effect of extinction). 
This only happens when the initial population is high enough to make low populations always highly improbable, \textit{i.e.}, $\mu(t) \gg \sigma(t) $ for all $t$. This requires $ x_0 + \Lambda t \gg \sqrt{2Dt} $ for all $t$, a condition that is not verified in our cases of interest (Table~\ref{tab:comparison}). 
Fig.~\ref{fig:pdf-comparison} shows that for middle times, this simple Gaussian approximation is not sufficient, {clearly overestimating the PDF}. We will address the necessary improvements in the next section, where we address the effects of extinction in reducing the norm and changing the shape of the PDF.

\begin{table}[h!]
\centering
% Aumenta la distancia entre columnas
\setlength{\tabcolsep}{3pt}
% Aumenta la distancia entre filas
\renewcommand{\arraystretch}{1.5}

\begin{tabular}{c|>{\centering\arraybackslash}p{2.4cm} >{\centering\arraybackslash}p{2.4cm} >{\centering\arraybackslash}p{2.4cm} >{\centering\arraybackslash}p{2.4cm}} 
\toprule
\makecell{Label} & \makecell{Maximum \\ growth (PF)} & \makecell{Minimum \\ extinction} & \makecell{Minimum \\ extinction (PF)} & \makecell{Low growth and \\ variance (PF)} \\ 
\midrule
$(\pi_{A\to B},\pi_{B\to A})$ & (0.263, 0.246) & (0.7, 0.1) & (0.569, 0.207) & (6.894, 0.001) \\ 
$\Lambda$ (collective) & 0.239 & 0.161 & 0.210 & $7.2\times10^{-5}$ \\ 
$\Lambda$ (agent, $n_0=40$) & 0.248 & 0.157 & 0.212 & 0.0056 \\ 
$\Lambda$ (agent, $n_0=10$) & 0.276 & 0.164 & 0.222 & 0.0065 \\ 
$D$ (collective) & 0.638 & 0.270 & 0.414 & 0.020 \\ 
$D$ (agent, $n_0=40$) & 0.570 & 0.279 & 0.387 & 0.0091 \\ 
$D$ (agent, $n_0=10$) & 0.548 & 0.276 & 0.365 & 0.0071 \\ 
\bottomrule
\end{tabular}
\caption{Comparison of growth $\Lambda$ and diffusion $D$ computed using the collective model approximation in Ref.~\cite{dinis_pareto-optimal_2022} and an agent-based model numerical simulation (as described in Sec.~\ref{sec:agentmodel}, until time $t=30.0$, $t=50.0$, $t=50.0$ and $t=1000.0$ respectively). {The rest of the parameters are fixed: $\kappa_{0\to 1} = \kappa_{1\to 0} = 1; k_{A0} = 2, k_{B0} = 0.2, k_{A1}= -2, k_{B1} = -0.2$.}}
\label{tab:comparison}
\end{table}

%%%%%%%%%%%%%%%%%%%%%%%%%%%%%%%%%%%%%%
\subsection{Extinction analysis} 
\label{sec:extinction}
\label{analizandoextincion}

We have focused our previous discussion on the growth and variance of the population.
Usually, variance is used as a proxy for extinction risk \cite{Gotelli2008,Lande2003a} (or bankruptcy risk in analogous financial systems \cite{Hull2005}). 
Here, we aim to explicitly compute and discuss the extinction risk of a population under bet-hedging and consider minimal extinction as another optimality criterion.  

The objective is to study the effect of bet-hedging. Consequently, we aim to determine whether reducing fluctuations {with respect to the maximum growth condition by} tuning the phenotype switching rates can result in a lower probability of extinction, even at the expense of instantaneous growth rate.

We first numerically compute the extinction probabilities in the agent-based model, a more accurate modelization of low populations and extinction phenomena. Afterward, we consider the collective model to assess the accuracy of the analytical expressions that it provides to estimate the extinction probability. 

%%%%%%%%%%%%%%%%%%%
\subsubsection{Extinction analysis in the agent-based model.} 
\label{sec:ext-agent-based}

Extinction plays a relevant role in the agent-based model dynamics, as already shown in Fig.~\ref{fig:pdf-comparison} {for the PDF of the (logarithm) of the population}. The agent-based model naturally presents extinction, as the dynamics can reach the extinction state, corresponding to a vanishing population in both phenotypes. Extinction is {an absorbing state} of the agent-based model dynamics, and the population cannot be recovered. 
 
\begin{figure}%[H] 
    \centering
    \subfigure[]{\includegraphics[draft=false,width=0.45\textwidth]{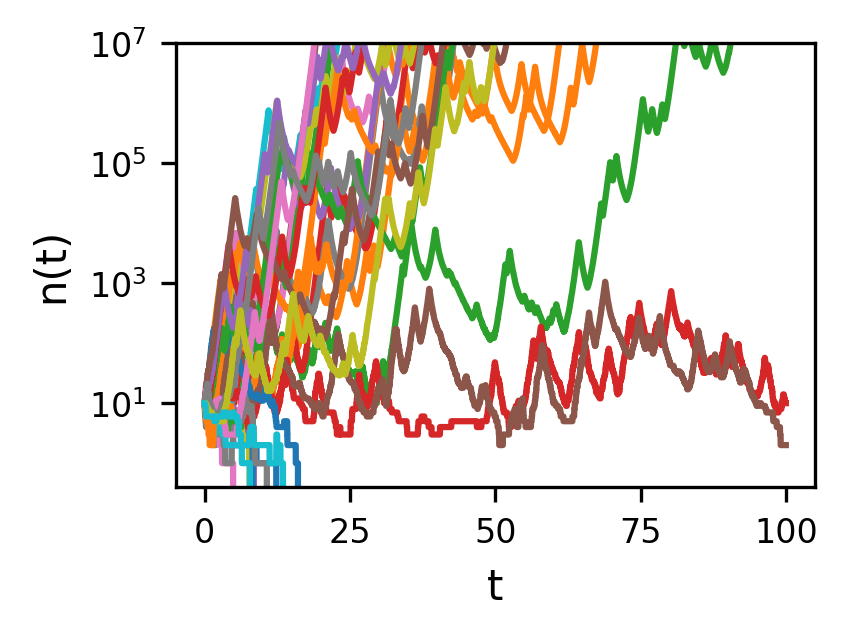}} 
    \subfigure[]{\includegraphics[draft=false,width=0.45\textwidth]{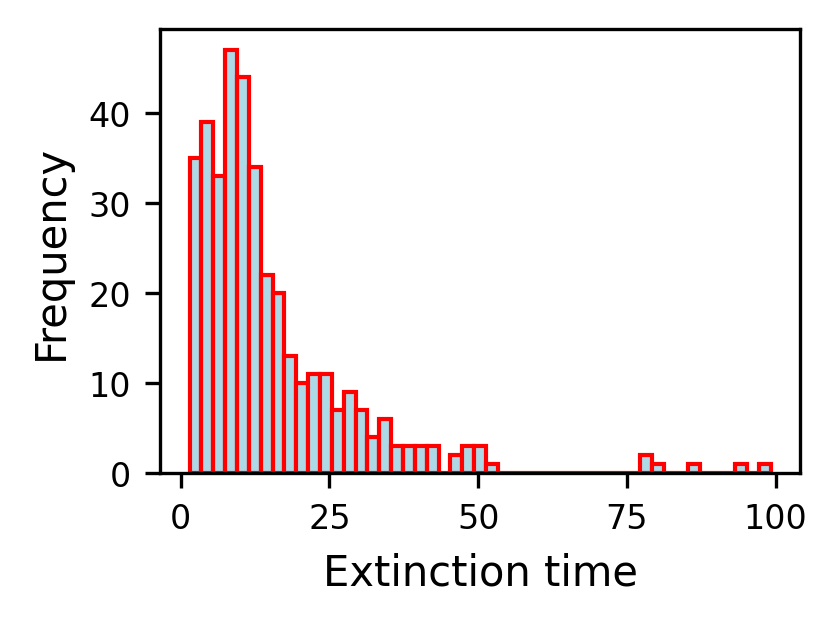}} 
    \caption{\textbf{Trajectories for the population and histogram of the extinction time} for the maximum growth (PF) case (whose switching rates are given at Table~\ref{tab:comparison}, the values of the other rates are given in the caption of Fig.~\ref{fig:populationdynamics}, with an initial population of $10$ individuals (5 of each phenotype)). The trajectories of $30$ simulations are represented, and the extinction time histogram is done with $1000$ simulations.}
    \label{tiempo300}
    \label{fig:trajectories+extinction-histogram}
\end{figure}

Extinction events occur mainly at early times when the population is low; see Fig.~\ref{fig:trajectories+extinction-histogram}. {At longer times, positive} mean growth moves the population away from extinction, reducing the extinction risk. For equal growth rates, a larger variability, given by the diffusion coefficient (or the variance), implies an increased extinction risk, mainly at early times. 

\begin{figure}%%[H] 
    \centering
    \subfigure[]{\includegraphics[draft=false,width=0.45\textwidth]{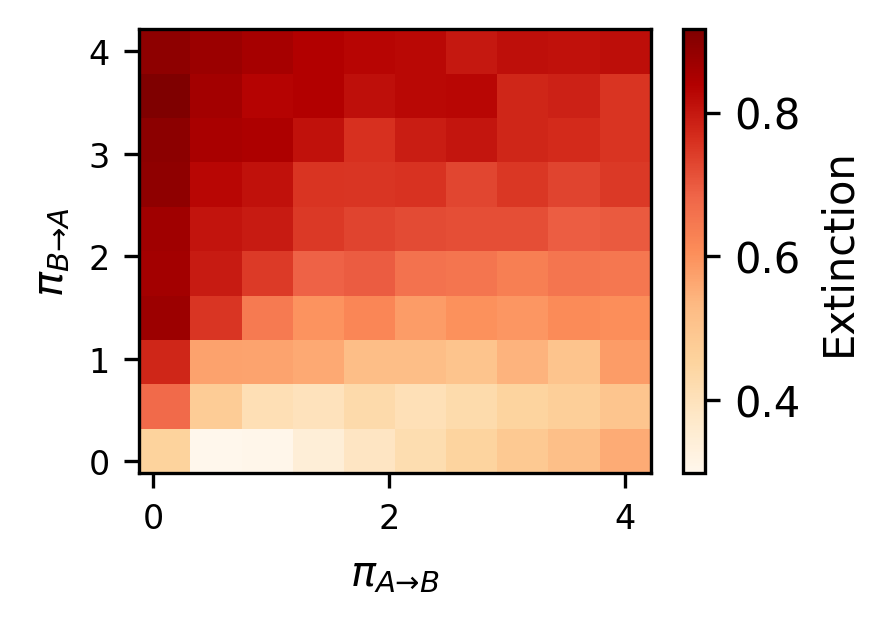}}
    \subfigure[]{\includegraphics[draft=false,width=0.45\textwidth]{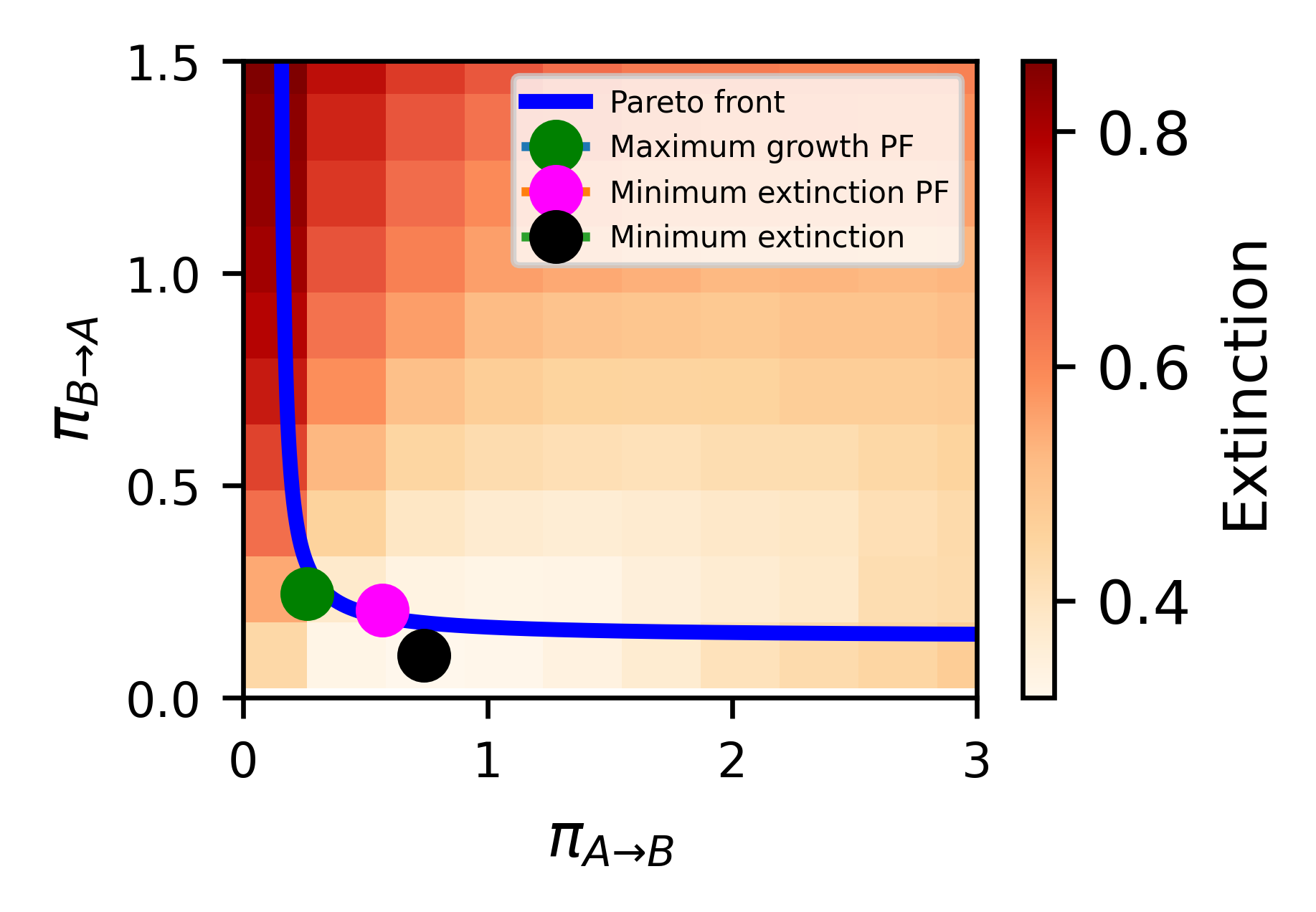}}
\caption{\textbf{Extinction probability} as a function of the switching rates (on a $10 \times 10$ mesh at time $1000$ with an initial population of $10$ individuals, $5$ of each phenotype, using for the other rates the values in the caption of Fig.~\ref{fig:populationdynamics}). $10^4$ simulations were done for each pair of phenotype switching values $\pi = (\pi_{A\to B}, \pi_{B\to A})$. {a) Scan of a wide region of phenotype switching rates where regions of lower probability are shown to be }concentrated around low values of $\pi_{B\to A}$ and up to $1.5$ for $\pi_{A\to B}$. { b) Computation in the low extinction region with a finer grid, depicting also the Pareto front.} The colored points correspond to the cases in Table \ref{tab:my-table}, where the {global} minimum extinction obtained in the simulations is shown in black, the minimum extinction inside the Pareto front in magenta, and the maximum growth from the Pareto front in green.}
\label{mapaextincion2}
\label{fig:extinction-probability-map}
\end{figure}

The extinction probability is defined as the fraction of realizations that lead to extinction (for a set of given fixed parameters). 
We sampled the extinction probability for a wide region of phenotype switching rate values, Fig.~\ref{fig:extinction-probability-map}, with the other parameters and initial conditions fixed to the values indicated in the caption of Fig.~\ref{fig:populationdynamics}. We have found that extinction presents a minimum  {near} the {collective's model} Pareto front {at a point in the parameter space given by the phenotype} switching rates values shown in Table~\ref{tab:extinction}. This switching rates imply a phenotype cycle time of $ 1/\pi_{A\to B} + 1/\pi_{B\to A} = 1/0.7 + 1/0.1 = 11.43 $, \textit{i.e.}, one order of magnitude greater than the environment cycle time $ 1/\kappa_{0\to 1} + 1/\kappa_{1\to 0} = 1 + 1 = 2 $. This {global} minimal extinction point presents a growth rate that is of the order of half the growth rate and half the dispersion coefficient of the maximum growth case; see Table~\ref{tab:comparison}. 

A similar extinction probability is also found for the point of minimum extinction in the Pareto front (see Table~\ref{tab:extinction} and Fig.~\ref{fig:extinction-probability-map}), whose switching cycle is $1/0.569 + 1/0.207 = 6.59 $, closer to the environment cycle time. This minimum extinction in the Pareto Front (PF) point has a higher growth rate, diffusion constant, and {a slightly higher} extinction probability; see Tables~\ref{tab:comparison} and \ref{tab:extinction}.

These results are compatible with the intuitive result from the collective model that reducing the fluctuations (diffusion coefficient), despite a (slight) reduction of the growth rate, can help reduce extinction. In the next subsection, we compare the collective model and the analytical results it provides.

Finally, we comment on a practical question for the numerical computation of the extinction probability with the agent-based model. Accurate computation requires computing many trajectory realizations over a long time. However, trajectories reaching a high population are unlikely to lead to later extinction. Thus, we fix an upper population limit that ends the computation and check that our extinction probability computation is independent of this upper limit, {once it is sufficiently large}. 

%(\textcolor{violet}{Pero esto que viene a continuación ya lo hemos comentado y no se añade nada, por qué lo repetimos aquí?)} Another practical question is the increasing number of individuals entering in the computation, as they are all equal an independent we can use the binomial distribution for the random change of the population of each phenotype. As we previously described in Section~\ref{sec:agentmodel}, this binomial dynamics is an exact result and does not introduce error. 

\begin{table}[h!]
\centering
% Aumenta la distancia entre columnas
\setlength{\tabcolsep}{3pt}
% Aumenta la distancia entre filas
\renewcommand{\arraystretch}{1.5}

\begin{tabular}{c|>{\centering\arraybackslash}p{2.4cm} >{\centering\arraybackslash}p{2.4cm} >{\centering\arraybackslash}p{2.4cm} >{\centering\arraybackslash}p{2.4cm}} 
\toprule
\makecell{Label} & \makecell{Maximum \\ growth (PF)} & \makecell{Minimum \\ extinction} & \makecell{Minimum \\ extinction (PF)} & \makecell{Low growth and \\ variance (PF)} \\ 
\midrule
$(\pi_{A\to B},\pi_{B\to A})$ & (0.263, 0.246) & (0.7, 0.1) & (0.569, 0.207) & (6.894, 0.001) \\ 
$\Lambda$ (collective) & 0.239 & 0.161 & 0.210 & $7.2\times10^{-5}$ \\
$D$ (collective) & 0.638 & 0.270 & 0.414 & 0.020 \\
$E$ (collective, $n_0=10$) & 0.4223 & 0.2511 & 0.3110 & 0.7128 \\ 
$E$ (agent, $n_0=10$) & 0.3670 & 0.2774 & 0.2959 & 0.8536 \\ 
$\epsilon_r(\%, n_0=10)$ & 13{.}09 & -10{.}47 & 4{.}85 & -19{.}75 \\
$E$ (collective, $n_0=40$) & 0.2513 & 0.1093 & 0.1540 & 0.5560 \\ 
$E$ (agent, $n_0=40$) & 0.1850 & 0.0906 & 0.1127 & 0.7062 \\ 
$\epsilon_r(\%, n_0=40)$ & 26{.}38 & 17{.}11 & 26{.}81 & -27{.}01 \\
\bottomrule
\end{tabular}
\caption{\textbf{Probability of extinction at time 1000 with initial population 10 (5 of each phenotype) and 40 (10 of each phenotype) for the same cases as Table~\ref{tab:comparison}.}
The theoretical extinction values were obtained using the collective approximation values of growth $\Lambda$, diffusion $D$, and extinction $E$ in Eq.~\ref{survivalfunctionused}. The simulated extinction value is calculated using the agent-based model and $10^4$  simulations. 
}
\label{tab:my-table}
\label{tab:extinction}
\end{table}

%%%%%%%%%%%%%%%%
\subsubsection{Extinction analysis in the collective model} \label{teorico} \label{sec:ext-collective}

The probability density function (PDF), Eq.~\ref{eq:pure_gaussian}, derived in Sec.~\ref{sec:collectivemodel} in the collective model for the logarithm of the total population number, $x=\ln(n)$, can also be obtained as the solution of the Fokker-Planck (FP) equation \cite{molini_first_2011}, 
\begin{equation}  \label{eq:FPeq}
    \frac{\partial p(x,t|x_0)}{\partial t} = - \Lambda \frac{\partial p(x,t|x_0)}{\partial x}+\frac{D}{2} \frac{\partial^2p(x,t|x_0)}{\partial x^2}
\end{equation}

where $p(x, t | x_0)$ is the transition probability density function (PDF), $x_0 = \ln(n_0)$ is the initial condition. $\Lambda$ and $D$ are the growth rate and diffusion coefficient, defined in Sec.~\ref{sec:collectivemodel}. They modulate the relevance of the drift and diffusion terms.

The extinction condition of the agent-based model $n=0$ cannot be directly translated to the collective model as it would give an absorbing condition at $x=\ln(0)=-\infty$ with no effect on the dynamics. A better approach is to compare the results of the agent-based model with extinction conditions at $n=0$ with the results of the collective model with extinction conditions at $n=1$, \textit{i.e.}, at $x=\ln(1)=0$. This implies solving the FP equation, Eq.~\ref{eq:FPeq}, with an absorbing boundary condition that reflects extinction given by $p(0, t) = 0$. 

Using the method of images \cite{molini_first_2011}, the explicit solution for this partial differential equation is

\begin{equation} \label{gaussinaconextincion}
p(x, t)=\frac{1}{\sqrt{4 \pi D t}}\left\{\exp \left[\frac{-\left(x-x_0-\Lambda t\right)^2}{4 D t}\right]-\exp \left(\frac{- \Lambda x_0}{D}\right) \exp \left[\frac{-\left(x+x_0-\Lambda t\right)^2}{4 D t}\right]\right\}.
\end{equation}
We are interested in finding the survival function $F(t | x_0)$, defined as the probability that extinction does not occur before time $t$ \cite{Cavallero2023.11.07.566039}. The extinction probability function is its complementary $E(t|x_0)=1-F(t | x_0)$. The survival function is calculated using the probability density function as
\begin{equation} 
    F(t|x_0)=\int^\infty_0p(x,t|x_0)dx.
\end{equation}
So, we arrive at the following result 
\begin{equation} \label{survivalfunctionused}
F\left(t \mid x_0\right) = \frac{1}{2} \left\{1-\mathrm{e}^{-\frac{\Lambda x_0}{D}}\left[1+\operatorname{erf}\left(\frac{\,{\Lambda}t-x_0}{\sqrt{4Dt}}\right)\right] +\operatorname{erf}\left(\frac{{\Lambda}t+x_0}{\sqrt{4Dt}}\right)\right\}.
\end{equation}
We should also remember that $x_0$ is the logarithm of the initial population, $x_0=\ln n(t=0)$. %Eq. \eqref{allows to give a theoretical prediction for the extinction probability E=1-F, using the analytical formulae for .... , see Table

%From $F$ we obtain the extinction probability $E=1-F$ and we use the values of $\Lambda$ and $ D $ obtained for the collective model in Ref.~\cite{dinis_pareto-optimal_2022} to study the different cases listed in table \ref{tab:extinction}.

%In this equation, we use the values of $\Lambda$ and $ D $ obtained in the collective model of each case studied. The analytical formulas for these constants, $\Lambda$ and $ D $, were obtained in Ref.~\cite{dinis_pareto-optimal_2022} and are used here. 

\begin{figure}%%[H] 
    \centering
    \subfigure[]{\includegraphics[draft=false,width=0.45\textwidth]{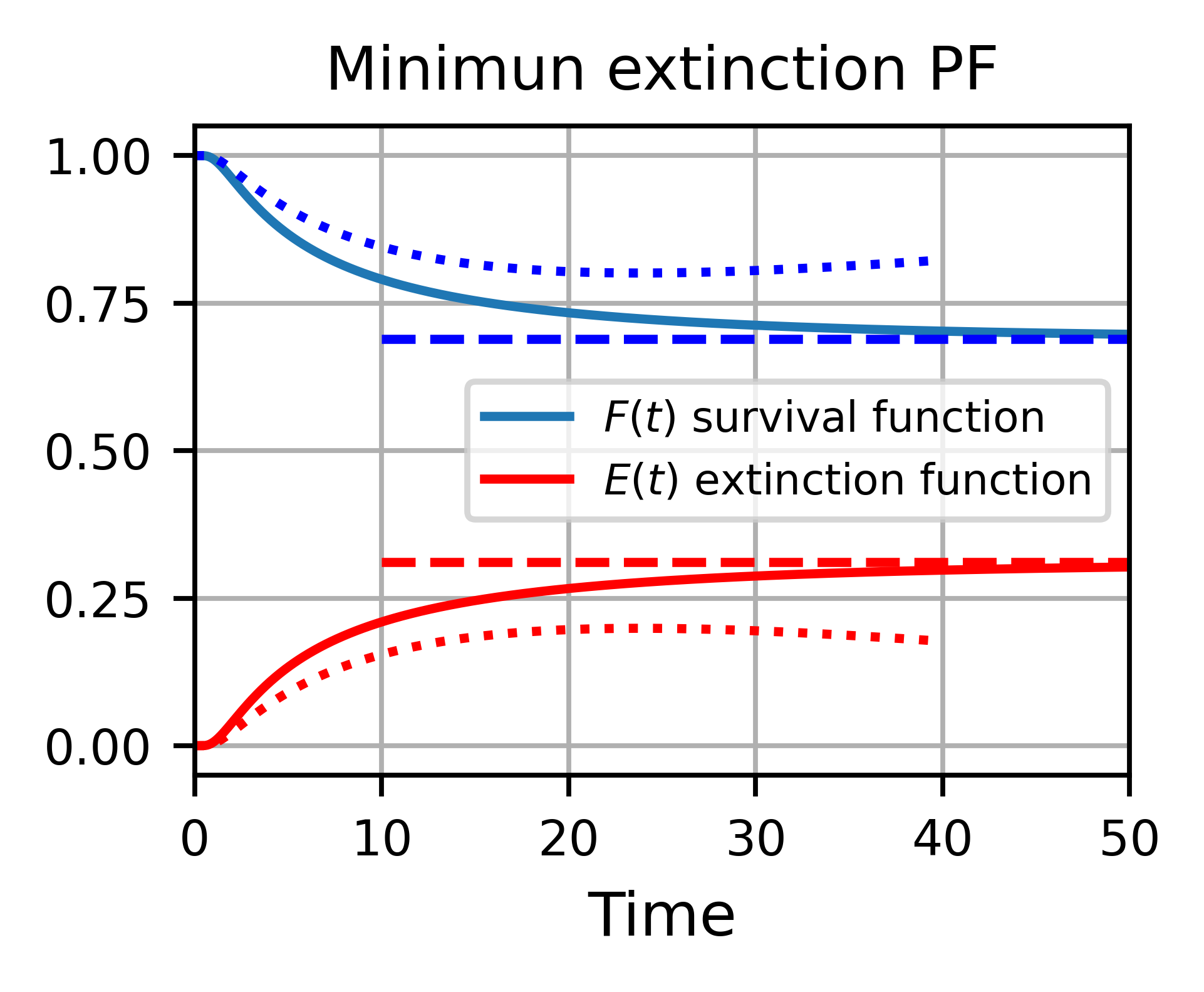}} 
    \subfigure[]{\includegraphics[draft=false,width=0.45\textwidth]{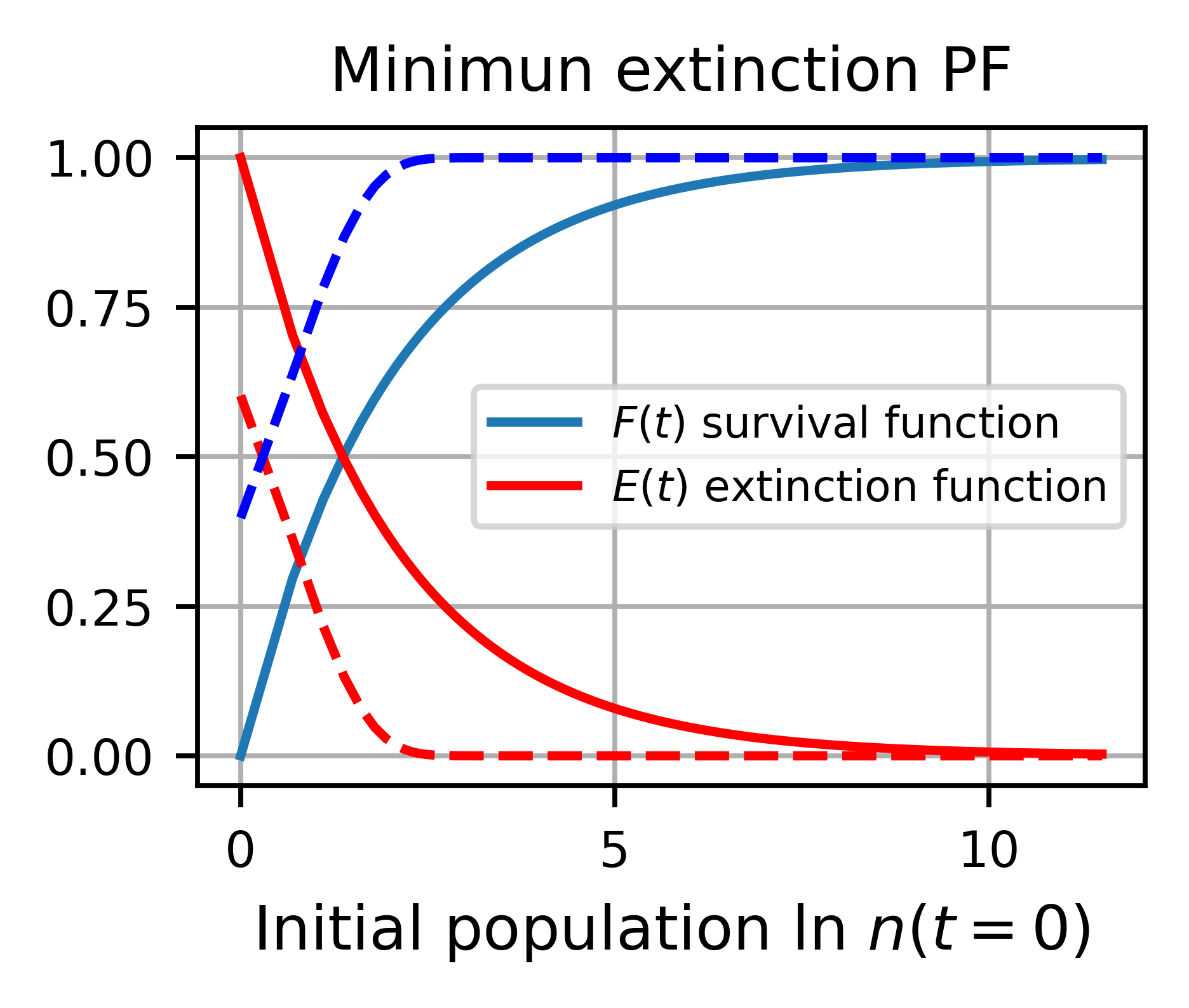}} 
    \caption{\textbf{Theorical survival and extinction probabilities} (solid lines) for the minimum extinction case in the Pareto Front (PF) (see Tables \ref{tab:comparison} and \ref{tab:extinction}) with initial population $n=10$, \textit{i.e.}, $x_0=\ln(10)$ using the $\Lambda$ and $D$ provided by the collective model, see Table~\ref{tab:comparison}. The dashed line shows the long-time approximation (Eq. \eqref{eq:extinction-longtime}), and the dotted line shows the short-time approximation (Eq. \eqref{eq:extinction-shorttime}).}
    \label{supervivencia}
    \label{fig:survival_extinction}
\end{figure}
The survival probability $F$ monotonously decreases over time, implying that extinction probability $E=1-F$ increases over time; see Fig.~\ref{fig:survival_extinction}a. 
For the minimum extinction in the Pareto front case, extinction nearly stops at times $ t \sim 50 $, where $ x = x_0 + \Lambda t \sim \ln(10) + 0.2 \cdot 50 = 10 $, which is a total population of $ n \sim \exp(10) = 2.2\cdot10^4$.
This is consistent with the observation, Fig.~\ref{fig:survival_extinction}b, that for this case, extinction is {practically} absent for initial populations with $x_0 = \ln(n_0) >10$, \textit{i.e.}, for initial populations $ n_0 > \exp(10) = 2.2\cdot10^4$. 

In the short time limit, we have for the extinction probability, $E=1-F$,
\begin{equation} \label{eq:extinction-shorttime}
    E(t|x_0) = 1 - F(t|x_0) = \sqrt{\frac{4Dt}{\pi x_0^2}} \; e^{\frac{-(x_0 + \Lambda t)^2}{4Dt}},
\end{equation}
while the asymptotic total extinction probability is given by
\begin{equation} 
\label{eq:extinction-longtime}
    E(t\to\infty | x_0 ) = 1 - F(t\to\infty | x_0 ) = e^{\frac{-\Lambda x_0}{D}}
    = (e^{-x_0})^\frac{\Lambda}{D} = \frac{1}{n_0^\frac{\Lambda}{D}}.
\end{equation}
These extinction probabilities computed with Eq. \eqref{survivalfunctionused} for the collective model, using the analytical formulae for $\Lambda$ and $D$ in Ref. \cite{dinis_pareto-optimal_2022}, are compared with the results previously obtained with the agent-based model in Table~\ref{tab:extinction}. The results show that the collective model underestimates the extinction probability, particularly in the cases with a smaller growth rate $\Lambda$. It also shows that the differences between the maximum growth rate case and the minimum extinction case for the agent-based model are smaller than predicted by the collective model.
The extinction expression, Eq.~\ref{eq:extinction-longtime}, is particularly enlightening. Note that for the maximum growth case (PF), we have $\frac{\Lambda}{D} \sim 0.37 (collective) $, while for the minimum extinction case, we have $\frac{\Lambda}{D} \sim 0.61 (collective) $ (if we consider instead the agent-based values the difference in the exponent reduces to $0.68$ vs. $0.80$). This implies that increasing the initial population by a factor of $ \alpha = n_0^{(0.61-0.37)/0.37} = 4$ is expected to have a similar effect to changing from the maximum growth strategy to the minimum extinction strategy (for the agent-based values the factor is $ \alpha = n_0^{(0.80-0.68)/0.68} = 1.4$ ).

\begin{figure}%[H]
    \centering
    \subfigure[]{\includegraphics[draft=false,width=0.45\textwidth]{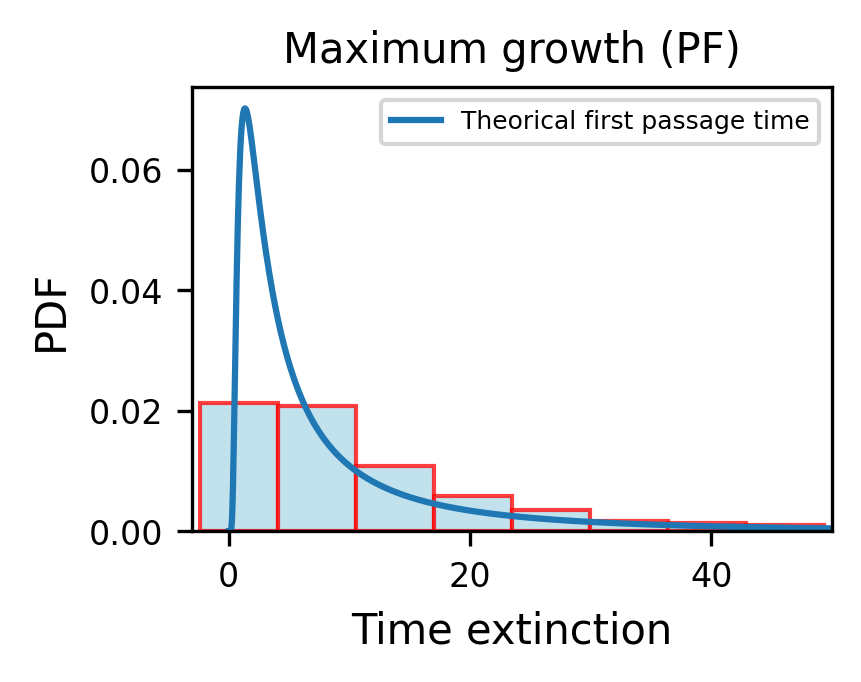}} 
    \subfigure[]{\includegraphics[draft=false,width=0.45\textwidth]{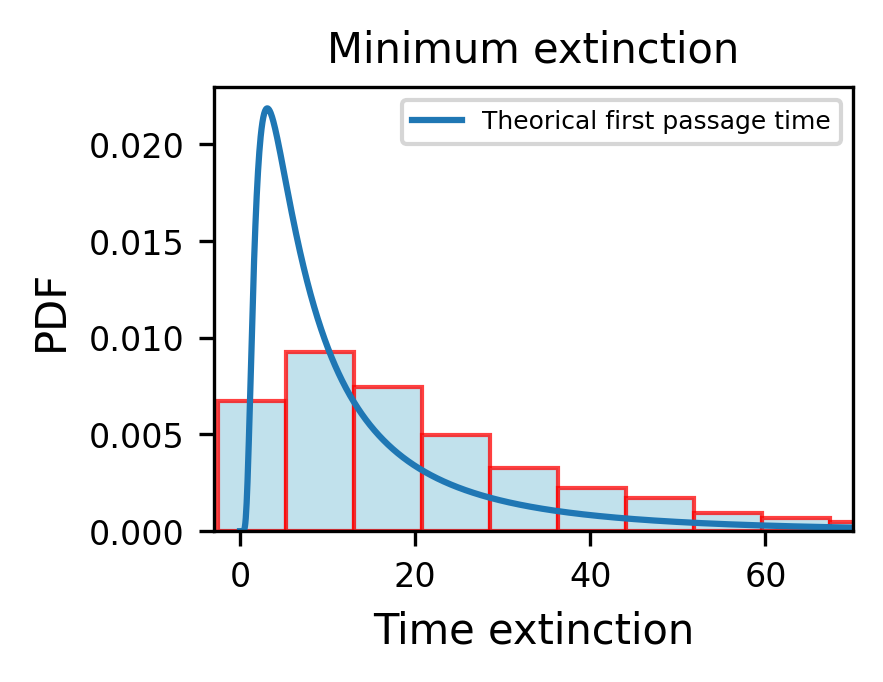}} 
	\subfigure[]{\includegraphics[draft=false,width=0.45\textwidth]{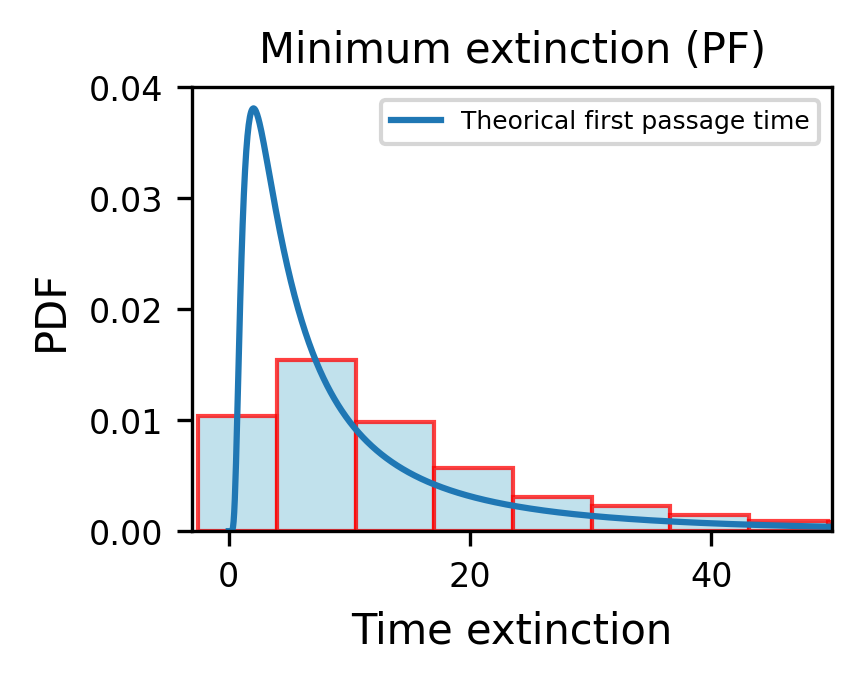}} 
    \subfigure[]{\includegraphics[draft=false,width=0.45\textwidth]{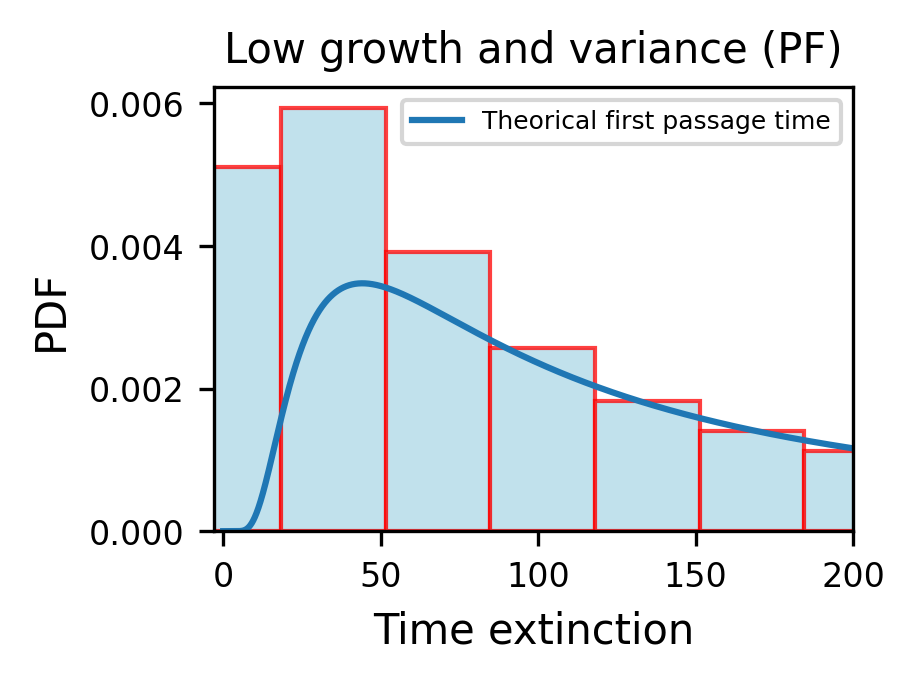}}
    \caption{\textbf{Extinction time probability density function.} Comparison of the collective model result (solid line), Eq.~\ref{eq:firstpassagetime}, and the agent-based model (bar plot), for the four cases in Tables~\ref{tab:comparison} and \ref{tab:extinction}. (Switching rates are given in the tables, the other rates and initial conditions are given in the caption of Fig.~\ref{fig:populationdynamics}.) 10000 simulations are performed for each case.
    }
    \label{fig:firstpassagetime}
    \label{primerapasada1}
\end{figure}

We can also compare the extinction time distribution function to get further insight into the comparison. The collective model extinction time distribution function is given by the \textbf{first-passage time function} \cite{molini_first_2011}, which is obtained as
\begin{equation} \label{eq:firstpassagetime}
g\left(t \mid x_0\right) = \frac{\partial}{\partial t} E(t\mid x_0) = \frac{x_0}{\sqrt{4 \pi D t^3}} \exp \left[\frac{-\left(x_0+\Lambda t\right)^2}{4Dt}\right] .
\end{equation}
Fig.~\ref{fig:firstpassagetime} shows a reasonable agreement between this theoretical extinction distribution function in the collective model, Eq.~\ref{eq:firstpassagetime}, and the numerical extinction time PDF obtained with numerical simulations of the agent-based model. The departure is more important for the cases with lower growth rate $\Lambda$, which are the minimum extinction case, and the low growth and variance at the PF case, in agreement with the comparison in Table~\ref{tab:extinction}. 

\newpage

%%%%%%%%%%%%%%%%%%%%%%%%%%%%%
%%%%%%%%%%%%%%%%%%%%%%%%%%%%%
\section{Conclusions}

We have shown with the agent-based model that the effects of finite population size are very relevant for the early-time dynamics of low populations. Collective model requirements of constant mean growth and diffusion coefficients are only reached for high populations, $100$ individuals or more. 

{
Our results corroborate that reducing fluctuations, even at the expense of a slight reduction of growth rate,  helps to reduce extinction in the vicinity of the maximum attainable growth rate. However, they also stress the role of the initial population. A situation where this may be relevant is for instance in the case of directed evolution in the lab, where many generations of microbes are evolved using a serial transfer protocol \cite{dettman_incipient_2007,paterson_antagonistic_2010}. In these experiments, a phase of growth is followed by dilution and a transfer to a fresh medium, repeating this cycle a number of times. 
Our results indicate that the population number just after the dilution, at the beginning of a new growth cycle may be determinant for extinction risk in the subsequent evolution. A large enough population greatly reduces the probability of extinction, making the actual strategy selected not strongly important. Conversely, if we would like to witness evolution of strategies sensitive to extinction-risk in the population, a significant dilution step where the population decreases to very low numbers would be desirable. In this case, finely tuning the strategy by the microbes may have a greater impact on their extinction risk than in cases where the population at the beginning of the growth cycle is large.
%For instance, maintaining a large enough initial population after a population bottleneck, as it occurs, for instance, in serial transfer experiments, is effective against extinction. Conversely, we can expect that to evolve extinction risk-sensitive strategies in a population, a significant dilution step where the population decreases to very low numbers would be needed. That would be the case where finely tuning the strategy may have a greater impact.
}
%\textcolor{red}{((Explicar serial transfer experiment. Cual es el bottleneck y  por que lo hay))((Varias de las frases, en particular la última me cuesta entenderlas))((enlazar con la siguiente frase/parrafo))((La idea puede ser que si no tenemos una población inicial grande, también es important el crecimiento rápido, ya que nos aleja de las poblaciones bajas)) Para enlazar, relacionar bottleneck con initial population number.}

%\textcolor{yellow}{However, it also stresses that keeping a high growth rate is crucial, as having a large population is more effective in preventing extinction than changing the phenotype switching rates. }
In this sense, the relevance of the population number is particularly stressed by Eq.~\ref{eq:extinction-longtime}, which states that the subsequent extinction after we reach a total population $n_0$ is of the order of $ 1/n_0^{\Lambda/D}$, where the exponent is in the interval between $1/3$ and $1$ for the maximum growth and the minimum extinction rate in the case considered here (see Fig.~\ref{fig:populationdynamics}). This reveals that a factor $4$ increase in the total population has a higher impact on reducing extinction than a change from the maximum growth strategy to the minimum extinction strategy.

Furthermore, the agent-based model also points out to the importance of demographic fluctuations for extinction. The collective model considered here included only a term equivalent to environmental noise, predicting %a lower extinction rate. 
different extinction probabilities.
Extensions of the collective model with demographic noise terms have for instance shown increased extinction probability \cite{Lande2003a}.

Our model included several simplifications. The most relevant one assumed that in the favourable environment, the dynamics of both phenotypes are dominated by reproduction (and death is neglected). In contrast, in the unfavourable environment, the dynamics of both phenotypes are dominated by death (and reproduction is neglected). Assuming that reproduction and death happen in both environments would imply an increase in the fluctuations, namely an increase in the diffusion coefficient. (This is a known result in the context of backward and forward molecular motors, see Ref.~\cite{Phillips2009}.)
We also neglected the genetic variability within the population.
A particularly interesting extension would consider evolving phenotypic change rates and their adaptability to changing conditions (such as changes in environment switching rates or the strength of the unfavourable environment). 
This would provide clues to understand the stability advantages of bet-hedging effects. 

\begin{acknowledgments}

LD and FJCG acknowledge financial support through Grants PID2020-113455GB-I00 and RTI2018-095802-B-I00, respectively, funded by the Ministerio de Ciencia e Innovaci\'{o}n (MINECO, Spain) and the European Regional Development Fund (ERDF).
The authors wish to thank David Lacoste for fruitful discussions and his careful reading of the manuscript.

\end{acknowledgments}

\bibliography{TFGFisica}

\end{document}